# Non-Markovian cost function for quantum error mitigation with Dirac Gamma matrices representation

Doyeol Ahn[1,*]

[1]Department of Electrical and Computer Engineering,
University of Seoul, 163 Seoulsiripdae-ro, Tongdaimoon-gu, Seoul 02504, Republic of Korea

*Corresponding author: dahn@uos.ac.kr

**ABSTRACT**

This paper investigates the non-Markovian cost function in quantum error mitigation (QEM) and employs Dirac Gamma matrices to illustrate two-qubit operators, significant in relativistic quantum mechanics. Amid the focus on error reduction in noisy intermediate-scale quantum (NISQ) devices, understanding non-Markovian noise, commonly found in solid-state quantum computers, is crucial. We propose a non-Markovian model for quantum state evolution and a corresponding QEM cost function, using simple harmonic oscillators as a proxy for environmental noise. Owing to their shared algebraic structure with two-qubit gate operators, Gamma matrices allow for enhanced analysis and manipulation of these operators. We evaluate the fluctuations of the output quantum state across various input states for identity and SWAP gate operations, and by comparing our findings with ion-trap and superconducting quantum computing systems' experimental data, we derive essential QEM cost function parameters. Our findings indicate a direct relationship between the quantum system's coupling strength with its environment and the QEM cost function. The research highlights non-Markovian models' importance in understanding quantum state evolution and assessing experimental outcomes from NISQ devices.

**Introduction**

Decoherence and gate errors are crucial noise sources in quantum computing systems, imposing practical constraints [1]. Yet, shallow quantum circuits, deployed in hybrid quantum-classical algorithms [2,3], can yield significant results, even amidst finite error rates, thanks to their error resilience [4,5]. Given the high cost of full-scale fault-tolerant quantum computing [6-8], it's vital to maximize error suppression while minimizing qubits, necessitating a solid understanding of the physical basis of these errors.

Current Quantum Error Mitigation (QEM) strategies involve gate models to describe quantum circuits and time-dependent Hamiltonian dynamics, employing the Lindblad operator [6] as a noise source [9].

A quantum system is considered as an open quantum system, coupled to an environment that induces decoherence and dissipation. The dynamics of an open quantum system can be described, in many cases, with a Markov approximation, which assumes that the environment recovers instantly from the interaction, leading to a continuous flow of information from the system to the environment. However, with the increasing capability of controlling quantum systems, a large separation between system and environment time scale can no longer be assumed, leading to non-Markovian behavior and eventually a back flow of information from the environment to the system. It is therefore crucial to develop an accurate but efficient description of the system-environment interaction that goes beyond the Markovian approximation [10-12].

While most studies focus on Markovian noise affecting Noisy Intermediate-Scale Quantum (NISQ) computers, it's essential to investigate non-Markovian noise sources, which exhibit memory effects and are prevalent in solid-state quantum computing devices [10-12]. Non-Markovian noise poses more complex modeling and error mitigation challenges, emphasizing the need for a deeper understanding for effective strategy implementation.

This research introduces a non-Markovian method for assessing quantum state fluctuations in near-term quantum computing devices to calculate the QEM cost function. Considering the ubiquity of non-Markovian noise effects in these devices, understanding their impact is crucial for effective error mitigation.

By employing this non-Markovian method and modeling the environment with simple harmonic oscillators, we can explore the correlation between noise sources and system performance. This is key to determining the QEM cost function, which measures the success of error reduction techniques in a quantum system.

The knowledge derived from this method aids in crafting custom QEM strategies, enabling researchers to address non-Markovian noise challenges more effectively. This leads to enhanced performance and reliability of quantum computing devices amidst various environmental disturbances.

Dirac Gamma matrices are a set of matrices that play a central role in the formulation of the relativistic quantum mechanics known as the Dirac equation, which describes the behavior of spin-1/2 particles such as electrons [13,14]. These matrices possess unique algebraic properties, including anticommutativity and a specific representation in terms of the Pauli matrices. The Gamma matrices form a Clifford algebra and are employed in various areas of physics, including quantum field theory and supersymmetry.

Most two-qubit quantum gate operations can be described by the SU(2) group which is a mathematical structure widely used in physics, especially in the context of quantum mechanics, where it describes spin-1/2 particles like electrons. It is a special case of the special unitary group of degree 2 and is isomorphic to the group of quaternions of norm 1. Expanding SU(2) by Dirac gamma matrices usually refers to expressing SU(2) generators in terms of the gamma matrices. The purpose of doing this is to relate the mathematical structures of spin more explicitly. This expansion allows one to interrelate the spin and internal symmetries and can lead to a richer understanding of the underlying physics. Also when the SU(2) algebra is represented using the Dirac gamma matrices, it can help to connect different areas of theoretical physics, revealing new insights into the relationships between spin, isospin, and the behavior of fermions.

The connection between two-qubit gate operators and Dirac Gamma matrices arises from their shared algebraic structure. By expressing the two-qubit gate operators using Dirac Gamma matrices, we can exploit the properties of Gamma matrices to analyze and manipulate the operators more effectively. This representation can offer valuable insights into the behavior of two-qubit gates, especially in the presence of noise and decoherence. Furthermore, leveraging the algebraic properties of Dirac Gamma matrices can facilitate the development of novel error mitigation techniques and robust quantum gate designs, ultimately enhancing the performance of quantum computers.

Dissipation in quantum dynamics of a two-state or qubit system is widely encountered in physics and chemistry, and its description remains a significant theoretical challenge, especially when considering non-Markovian processes [10-12, 15]. These non-Markovian processes related to dissipation are closely linked to the low-frequency noise spectrum [16-18]. It has been suggested that low-frequency noise plays a crucial role in decoherence processes in ion trap or superconducting systems for near-term quantum computers [3]. Understanding non-Markovian quantum dynamics of a two-state system is also important for evaluating the performance of an adiabatic quantum computer in the presence of noise [19]. This is because, for many difficult problems, the computation bottleneck involves passing through a point where the gap between the ground state and the first excited state is small, and the system is strongly coupled to the environment. One of the authors previously developed a time-convolutionless reduced-density operator theory of noisy quantum-state evolution [20, 21]. This time-convolutionless formulation was shown to incorporate both non-Markovian relaxation and memory effects [22]. The time-convolutionless equations of motion were initially suggested by Tokuyama and Mori in the Heisenberg picture [23], and later developed in the Schrödinger picture using the projection operator technique [24-25].

In this work, we refine the reduced-density operator theory to create an adaptable expression for the quantum Liouville equation for a two-state system. This makes it more amenable to perturbational analysis, facilitating the derivation of the master equation in the context of low-frequency noise and strong environmental coupling.

This approach enables a deeper exploration of the impact of non-Markovian noise on quantum computing devices. By understanding how diverse noise sources affect system dynamics, researchers can pinpoint major error contributors and devise effective mitigation strategies.

The study of non-Markovian noise sources is pivotal to advancing quantum computing technologies. A non-Markovian approach to analyzing QEM cost function from the Dirac Gamma matrices representation of a recovery operation can lead to tailored error mitigation strategies, enhancing the performance and reliability of near-term quantum computing devices. This work will contribute to harnessing quantum computing's potential across various applications, including cryptography and materials science.

**Theoretical formulation**

The total Hamiltonian for an open two-state system is given by [20,21, 27]

$$\widehat{H}_T = \widehat{H}_S(t) + \widehat{H}_B + \widehat{H}_{int} \qquad (1)$$

where $\widehat{H}_S(t)$ is the system Hamiltonian for a two-state system, $\widehat{H}_B$ the Hamiltonian acting on the reservoir or an environment, $\widehat{H}_{int}$ is the interaction between the system and the environment (as depicted in figure 1).

The equation of motion for the total density operator $\rho_T(t)$ of the total system is given by a quantum Liouville equation [20]

$$\frac{d}{dt}\rho_T(t) = -i[\widehat{H}_T(t), \rho_T(t)] = -i\widehat{L}_T(t)\rho_T(t), \qquad (2)$$

where

$$\widehat{L}_T(t) = \widehat{L}_S(t) + \widehat{L}_B(t) + \widehat{L}_{int}(t) \qquad (3)$$

is the Liouville super operator in one-to-one correspondence with the Hamiltonian. Here, we use the unit in which $\hbar = 1$. In order to derive an equation and to solve for a system alone, it is convenient to use the projection operators which decompose the total system by eliminating the degrees of freedom for the reservoir. We define thine-independent projection operator $\underline{P}$ and $\underline{Q}$ given by [20]

$$\underline{P}X = \rho_B tr_B X, \underline{Q} = 1 - \underline{P} \tag{4}$$

for any dynamical variable $X$. Here $tr_B$ denotes a partial trace over the quantum reservoir and $\rho_B$ is the density matrix of the reservoir.

The information of the system is then contained in the reduced density operator

$$\rho(t) = tr_B \rho_T(t) = tr_B \underline{P} \rho_T(t). \tag{5}$$

We assume that the system was turned on at $t = 0$ and the input state prepared at $t = 0$ was isolated with the reservoir such that $\underline{Q}\rho_T(0) = 0$.

The formal solution of Eq. (2) is given by (Supplementary Information)

$$\rho_B \rho(t) = \rho_B \widehat{U}_S(t,0)\rho(0) - i\rho_B \int_0^t ds \widehat{U}_S(t,s) tr_B \left[\widehat{L}_T(s)\{\widehat{\theta}(s) - 1\}\rho_B\right] tr_B \left[\underline{G}(t,s)\widehat{\theta}(t)\rho_B\right] \rho(t), \tag{6}$$

or

$$\left(1 + i\int_0^t ds \widehat{U}_S(t,s) tr_B \left[\widehat{L}_T(s)\{\widehat{\theta}(s) - 1\}\rho_B\right] tr_B \left[\underline{G}(t,s)\widehat{\theta}(t)\rho_B\right]\right)\rho(t) \tag{7}$$
$$= \widehat{U}_S(t,s)\rho(t).$$

Here $\widehat{U}_S(t,s)$ is the propagator of the system defined by

$$\widehat{U}(t,s)\underline{P} = \underline{T}exp\left[-i\int_s^t d\tau \underline{P}\widehat{L}_T(\tau)\underline{P}\right]\underline{P}$$
$$= \underline{T}exp\left[-i\int_s^t d\tau \widehat{L}_S(\tau)\underline{P}\right]\underline{P} \tag{8}$$
$$= \underline{T}exp\left[-i\int_s^t d\tau \widehat{L}_S(\tau)\right]\underline{P} = \widehat{U}_S(t,s)\underline{P}.$$

If we define $\widehat{W}(t)$ by

$$\widehat{W}(t) = 1 + i\int_0^t ds \widehat{U}_S(t,s) tr_B \left[\widehat{L}_T(s)\{\widehat{\theta}(s) - 1\}\rho_B\right] tr_B \left[\underline{G}(t,s)\widehat{\theta}(t)\rho_B\right], \tag{9}$$

Then, the evolution operator for the reduced density operator $\rho(t)$ is given by

$$\rho(t) = \widehat{W}^{-1}(t)\widehat{U}_S(t,0)\rho(0) = \widehat{V}(t)\rho(0), \tag{10}$$

where the super-operator $\widehat{V}(t)$ for the evolution of the reduced density operator is defined by $\widehat{V}(t) = \widehat{W}^{-1}(t)\widehat{U}_S(t,0)$. Within the Born approximation, we have (Supplementary Information)

$$\widehat{V}^{(2)}(t) = \left\{1 - \int_0^t ds \int_0^s d\tau \widehat{U}_S(t,s) tr_B \left[\widehat{L}_{int}\widehat{U}_o(s,\tau)\widehat{L}_{int}\widehat{U}_o^{-1}(s,\tau)\rho_B\right]\widehat{U}_o^{-1}(t,s)\right\}\widehat{U}_S(t,0), \tag{11}$$

where

$$\widehat{U}_o(t) = \underline{T}exp\left[-i\int_0^t d\tau \left(\widehat{L}_S(\tau) + \widehat{L}_B\right)\right]$$
$$= exp(-it\widehat{L}_B)\underline{T}exp\left[-i\int_0^t d\tau \left(\widehat{L}_S(\tau)\right)\right] \quad (12)$$
$$= \widehat{U}_B(t)\widehat{U}_S(t).$$

After some mathematical manipulations, $\widehat{V}^{(2)}(t)\rho(0)$ becomes

$$\rho(t)$$
$$= \widehat{V}^{(2)}(t)\rho(0)$$
$$= \widehat{U}_S(t)\rho(0) - \int_0^t ds \int_0^s d\tau\, tr_B\left[\widehat{H}_{int}\widehat{H}_{int}(\tau-s)\rho_B\rho(-s)\right] + \int_0^t ds \int_0^s d\tau\, tr_B\left[\widehat{H}_{int}(\tau-s)\rho_B\rho(-s)\rho_B\widehat{H}_{int}\right]$$
$$+ \int_0^t ds \int_0^s d\tau\, tr_B\left[\widehat{H}_{int}\rho_B\rho(-s)\widehat{H}_{int}(\tau-s)\right] - \int_0^t ds \int_0^s d\tau\, tr_B\left[\rho_B\rho(-s)\widehat{H}_{int}(\tau-s)\widehat{H}_{int}\right]. \quad (13)$$

Here $\widehat{H}_{int}(\tau-s)$ and $\rho(-s)$ are Heisenberg operators defined by

$$\widehat{H}_{int}(\tau-s) = exp\left(i\widehat{H}_o(\tau-s)\right)\widehat{H}_{int}exp\left(-i\widehat{H}_o(\tau-s)\right),$$
$$\rho(-s) = exp\left(-i\int_0^s dt\widehat{H}_S(t)\right)\rho(0) exp\left(i\int_0^s dt\widehat{H}_S(t)\right),$$

respectively.

**Results**

In this work, we focus on the two-qubit gate operations and model the interaction of the quantum system with the environment during the gate operation by a Caldeira-Leggett model [11-12,15, 27] where a set of harmonic oscillators are coupled linearly with the system spin by

$$\widehat{H}_{int} = \lambda \sum_{i=1,2,\, j=1,2,3} \vec{S}_i \cdot \vec{b}_i,\quad S_i^j = \frac{\hbar}{2}\sigma_i^j \quad (14a)$$

where $\sigma_i^j$ are the Pauli matrices $\sigma_i^1 = X_i,\ \sigma_i^2 = Y_i,\ \sigma_i^3 = Z_i$ and $b_i^j$ is the fluctuating quantum field associated the ith qubit, whose motion is governed by the harmonic-oscillator Hamiltonian. We model the system by the Hubbard Hamiltonian

$$H_S(t) = J(t)\vec{S}_1 \cdot \vec{S}_2 \quad (14b)$$

where $J(t)$ is the time-dependent Heisenberg coupling for electron spin operators $\vec{S}_1,\ \vec{S}_2$.

If we turn on $J(t)$ for $\int dt\, J(t) = \pi$, the unitary operator associated with Eq. (14b) gives the SWAP operation up to the overall phase [20, 27]

In the evaluation of Eq. (13), we obtain the following relations [20]:

$$tr_B\{b_k^l(t) b_i^j \rho_B\} = \delta_{ik}\delta_{jl}[\Gamma(t) + i\Delta(t)],$$
$$tr_B\{b_i^j b_k^l(t) \rho_B\} = \delta_{ik}\delta_{jl}[\Gamma(t) - i\Delta(t)],$$
(15)

where

$$\Gamma(t) + i\Delta(t) = \frac{\lambda^2}{\pi}\int_0^\infty J(\omega)\left\{\exp(-i\omega t) + \coth\left(\frac{\omega}{2k_B T}\right)\cos\omega t\right\}.$$
(16)

Here $J(\omega)$ is the ohmic damping given by $J(\omega) = \theta(\omega_c - \omega)\eta\omega$, $\Gamma$ is the decoherence rate of the qubit system, $\eta$ is $\frac{\omega_c}{T}$ and $\omega_c$ is the cutoff frequency.

**1) Evaluation of a reduced-density-operator in the multiplet basis representation**

We evaluate the reduced-density-operator in the multiplet basis representation [20]

$$\rho(t) = \sum_{a,b}\rho_{ab}(t) e_{ab}, \quad e_{ab} = |a><b|, \quad a,b = 1,2,3,4,$$
(17)

where $e_{ab}$ is the multiplet states (basis). The inner product between the multiplet basis is defined by

$$(e_{ab}, e_{cd}) = tr[e_{ab}^\dagger e_{cd}] = \delta_{ac}\delta_{bd}.$$
(18)

Then, from Eqs. (16)-(21), we obtain the matrix component of the reduced-density-operator as (Supplementary Information)

$$\rho_{ab}(t) = V_{ab|cd}(t)\rho_{cd}(0),$$
$$V_{ab|cd}(t) = \exp[-it(E_a - E_b)]\left\{\delta_{ac}\delta_{bd} - \left[\delta_{bd}\sum_{a'}M_{aa'a'c} - M_{acdb}\right]k(t) - \left[\delta_{ac}\sum_{a'}M_{aa'a'b} - M_{acdb}\right]k^*(t)\right\}$$
(19)

where

$$M_{abcd} = \sum_{i,j}<a|S_i^j|b><c|S_i^j|d>$$
$$= \frac{1}{4}\sum_{i=1,2}\{<a|X_i|b><c|X_i|d> + <a|Y_i|b><c|Y_i|d> + <a|Z_i|b><c|Z_i|d>\},$$
(20)

and

$$k(t) = \frac{2}{\pi}\Gamma_o\left\{\frac{\pi}{2}\omega_c t + \int_0^{\frac{t}{\tau_s}} Si(\omega_c\tau_s t) dt\right\}$$
$$+ i\Delta_o\left\{\frac{t}{\tau_s} - \frac{1}{\omega_c\tau_s}\left(\frac{\pi}{2}\omega_c\tau_s + Si(\omega_c t)\right)\right\}.$$
(21)

Here $\omega_c$ is the high frequency cutoff, $\tau$ is the switching time, $\Gamma_0 = \lambda^2\eta k_B T\tau_s$ and $\Delta_o = \frac{\lambda^2\eta\omega_c\tau_s}{\pi}$. We now study the non-Markovian errors associated with two-qubit gate operations. In Eq. (21), $Si(x) = \int_0^x dt\frac{\sin t}{t}$, is a sine integral [29].

In Eq. (19), $V_{ab|cd}(t)$ is the quantum evolution operator for the reduced density operator containing the effects of non-Markovian noise sources. The quantum evolution operator $V_{ab|cd}(t)$ is defined in the multiplet basis $e_{ab} = |a\rangle\langle b|$ by $V_{ab|cd}(t) = (e_{ab}, V(t) e_{cd}) = Tr(e_{ab}^\dagger V(t) e_{cd})$ where $V(t)$ is deined by Equation (10). In Equation (19), $E_a, E_b$ are the steady-state eigenvalues (energy) of the system Hamiltonian $H_S(t)$ for the states $|a\rangle, |b\rangle$, respectively.

The interaction between the qubit system and the environment defined in Eq. (15) is assumed to be a local interaction. When qubits are entangled one may need to consider the non-local interactions as well. Recent experiments have demonstrated significant non-local noise across different qubits in the spin qubit platforms [30-32] and relevant theoretical exploration [33].

## 2) Schematics of quantum error mitigation process

The purpose of QEM is to restore the ideal quantum evolution without noisy processes. We denote the ideal quantum volution and QEM based recovery operation as $\varepsilon_I(t)$ and $R_{QEM}(t)$, respectively, such that [6,10]

$$\varepsilon_{I\,ab|cd}(t) = \sum_{e,f} R_{QEM\,ab|ef}(t) V_{ef|cd}(t), \qquad (22)$$

and $\quad R_{QEM}(t) = \sum_i \mu_i O_i = c(t) \sum_i sgn(\mu_i) p_i O_i. \qquad (23)$

Here, $c(t) = \sum_i |\mu_i|$ is the QEM cost function, $p_i = \dfrac{|\mu_i|}{c(t)}$, and $\{O_i\}$ is the set of physical operations applied for QEM.

Equation (22) can be rewritten as

$$\varepsilon_I(t) = R_{QEM}(t) V(t) \text{ or } R_{QEM}(t) = \varepsilon_I(t) V^{-1}(t).$$

Therefore, we can interpret $R_{QEM}(t)$ as a recovery operator and $\mu_i$ is the expansion coefficient of a recovery operator for given basis i such as Dirac matrices and can be interpreted as quasi-probability. As a result, $c(t)$ can be interpreted as probabilistic error cancellation or cost function.

## 3) Non-Markovian noise of SWAP operation

We first consider the SWAP operation.

For SWAP operation, the multiple basis is given by

$$|1\rangle^m = |00\rangle,$$
$$|2\rangle^m = \frac{1}{\sqrt{2}}(|01\rangle + |10\rangle),$$
$$|3\rangle^m = |11\rangle, \qquad (24)$$
$$|4\rangle^m = \frac{1}{\sqrt{2}}(|01\rangle - |10\rangle)$$

Assuming that the initial state is given by $\rho_{cd}(0) = |00\rangle\langle 00| = \rho_{11}^m(0)$, we obtain

$$\begin{aligned}
\rho_{11}^m(t) &= V_{11|11}(t)\rho_{11}^m(0) \\
&= \left\{1 - 2\left[\sum_{a'} M_{1a'a'1} - M_{1111}\right] Re\, k(t)\right\} \rho_{11}^m(0) \\
&= (1 - 2\, Re\, k(t))\rho_{11}^m(0), \\
\rho_{22}^m(t) &= V_{22|11}(t)\rho_{11}^m(0) \\
&= 2M_{2112}\, Re\, k(t)\, \rho_{11}^m(0) \\
&= Re\, k(t)\, \rho_{11}^m(0), \\
\rho_{33}^m(t) &= V_{33|11}(t)\rho_{11}^m(0) \\
&= 2M_{3113}\, Re\, k(t)\, \rho_{11}^m(0) \\
&= 0, \\
\rho_{44}^m(t) &= V_{44|11}(t)\rho_{11}^m(0) \\
&= 2M_{4114}\, Re\, k(t)\, \rho_{11}^m(0) \\
&= Re\, k(t)\, \rho_{11}^m(0).
\end{aligned} \qquad (25)$$

In Eq. (25), $M_{2112}$ is, for example, given by

$$\begin{aligned}
M_{2112} = &\frac{1}{4}\langle 2|X_1|1\rangle\langle 1|X_1|2\rangle + \frac{1}{4}\langle 2|Y_1|1\rangle\langle 1|Y_1|2\rangle + \frac{1}{4}\langle 2|Z_1|1\rangle\langle 1|XZ_1|2\rangle \\
&+ \frac{1}{4}\langle 2|X_2|1\rangle\langle 1|X_2|2\rangle + \frac{1}{4}\langle 2|Y_2|1\rangle\langle 1|Y_2|2\rangle + \frac{1}{4}\langle 2|Z_2|1\rangle\langle 1|Z_2|2\rangle = \frac{1}{2}
\end{aligned} \qquad (26)$$

from Eq. (20).

In NISQ machines, the input and out states are represented by the computational basis:

$$e_{\alpha\beta}^c = |\alpha\rangle\langle\beta|,\ \alpha,\beta = 1,2,3,4,\ \{|00\rangle,|01\rangle,|10\rangle,|11\rangle\}. \qquad (27)$$

Then the reduce-density-operator in the computational basis is given by

$$\begin{aligned}
\rho_{\alpha\beta}^c(t) &= \sum_{a',b'} C_{\alpha\beta|a'b'}\, \rho_{a'b'}^m(t), \\
C_{\alpha\beta|ab} &= tr\left(e_{\alpha\beta}^{c\dagger}, e_{ab}^m\right).
\end{aligned} \qquad (28)$$

By substituting, Eqs. (27). (28) into Eq. (25), we obtain '

$$\begin{aligned}
\rho_{11}^c(t) &= (1 - 2\, Re\, k(t))\, \rho_{11}^m(0), \\
\rho_{22}^c(t) &= Re\, k(t)\, \rho_{11}^m(0), \\
\rho_{33}^c(t) &= Re\, k(t)\, \rho_{11}^m(0) \\
\rho_{44}^c(t) &= 0.
\end{aligned} \qquad (29)$$

Here $1 - 2Re\, k(t)$, $Re\, k(t)$, $Re\, k(t)$, $0$ are the probabilities of finding the output states of the system in $|00\rangle, |01\rangle, |10\rangle, |11\rangle$ states at time $t$, respectively.

It is interesting to note that the $Re\, k(t)$ can be approximated as

$$Re\ k(t) \approx \frac{2}{\pi}\Gamma_0\omega_c\tau_s\left[\frac{\pi}{2}\frac{t}{\tau_s}+\frac{1}{2}\left(\frac{t}{\tau_s}\right)^2\right]. \tag{30}$$

In figure 2, we show the plot of $Re\ k(t)$ vs $\frac{t}{\tau_s}$, where $\tau_s$ is the switching time for the parameters $\Gamma_0\omega_c\tau_s$ in the range $7.0\times 10^{-4} \leq \Gamma_0\omega_c\tau_s \leq 7.0\times 10^{-3}$.

The comprehensive summary provided in Table 1 encapsulates the theoretical forecasts concerning fluctuations in quantum states that are inherent in the execution of the SWAP operation, particularly when these operations are subject to non-Markovian noise sources. These predictions, drawn from theoretical work, detail the dynamic behaviors and potential variations of quantum states when a SWAP operation - a fundamental quantum gate operation used to exchange the quantum states of two qubits - is applied. The aspect of non-Markovian noise underlines the consideration of more complex noise patterns, where the system's future states depend not just on its present state, but also on its past, introducing memory effects and making the noise process more intricate and challenging to manage.

Table 1: Quantum state fluctuations of SWAP operation with non-Markovian noise sources

|  |  | Input state | | | |
|---|---|---|---|---|---|
|  |  | $\|1\rangle^m$ | $\|2\rangle^m$ | $\|3\rangle^m$ | $\|4\rangle^m$ |
| Output state | $\|00\rangle$ | $1-2Re\ k(t)$ | $Re\ k(t)$ | 0 | $Re\ k(t)$ |
|  | $\|01\rangle$ | $Re\ k(t)$ | $\frac{1-2Re\ k(t)}{2}$ | $Re\ k(t)$ | $\frac{1-2Re\ k(t)}{2}$ |
|  | $\|10\rangle$ | $Re\ k(t)$ | $\frac{1-2Re\ k(t)}{2}$ | $Re\ k(t)$ | $\frac{1-2Re\ k(t)}{2}$ |
|  | $\|11\rangle$ | 0 | $Re\ k(t)$ | $1-2Re\ k(t)$ | $Re\ k(t)$ |

In an effort to gauge the fluctuations in quantum states during the execution of the SWAP operation on practical NISQ devices, two distinct platforms were utilized: 'ibm_guadalupe' via IBM Quantum, and IonQ facilitated by Amazon Braket. The data derived from these platforms offer critical insights into the behavior of quantum states under the SWAP operation. After conducting 1,000 iterations on each platform, the resulting probabilities of each quantum state have been meticulously recorded and presented in Table 2 for 'ibm_guadalupe' and Table 3 for IonQ. These tables provide a comparative analysis of the two platforms, aiding in understanding the influence of different NISQ devices on quantum state fluctuations during the SWAP operation.

Table 2: Quantum state fluctuations under the SWAP operation for IonQ

|  |  | Input state | | | |
| --- | --- | --- | --- | --- | --- |
|  |  | $|1>^m$ | $|2>^m$ | $|3>^m$ | $|4>^m$ |
| Output state | $|00>$ | 0.955 | 0.023 | 0.005 | 0.023 |
|  | $|01>$ | 0.017 | 0.518 | 0.012 | 0.474 |
|  | $|10>$ | 0.018 | 0.453 | 0.011 | 0.493 |
|  | $|11>$ | 0.010 | 0.006 | 0.0972 | 0.010 |

Table 3: Quantum state fluctuations under the SWAP operation for ibm_guadalupe

|  |  | Input state | | | |
| --- | --- | --- | --- | --- | --- |
|  |  | $|1>^m$ | $|2>^m$ | $|3>^m$ | $|4>^m$ |
| Output state | $|00>$ | 0.936 | 0.056 | 0.015 | 0.040 |
|  | $|01>$ | 0.020 | 0.404 | 0.033 | 0.410 |
|  | $|10>$ | 0.024 | 0.524 | 0.025 | 0.532 |
|  | $|11>$ | 0.020 | 0.016 | 0.927 | 0.016 |

From the above table, the decoherence function $Re\ k(t)$ at switching time $\tau_s$ can be estimated for each NISQ machine as in the range $6.0 \times 10^{-3} \leqslant Re\ k(\tau_s^{IonQ}) \leq 1.8 \times 10^{-2}$ for IonQ, and $1.5 \times 10^{-2} \leqslant Re\ k(\tau_s^{IBM}) \leq 5.6 \times 10^{-2}$ for ibm-guadrupe, respectively.

**4) Non-Markovian noise of an Identity operation**

For identity operation, the input and out states are represented by the computational basis:

$$e_{\alpha\beta}^c = |\alpha><\beta|,\ \alpha,\beta=1,2,3,4,\ \{|00>,|01>,|10>,|11>\}.$$

Table 4 offers an encapsulation of theoretical projections concerning the fluctuations in quantum states during the execution of an Identity operation, an essential quantum gate operation. Contrary to the previous section, the Identity operation does not interchange the states of two qubits, but rather leaves the state of a qubit unchanged. This becomes increasingly crucial when these operations are impacted by non-Markovian noise sources. These predictions explore the possible alterations and dynamic responses of quantum states during the Identity operation. Non-Markovian noise, representing a more intricate noise pattern, highlights the system's dependence not merely on its current state, but also its past states. This inclusion of historical states introduces memory effects into the system, subsequently complicating the noise process and intensifying the challenge of managing it.

Table 4: Quantum state fluctuations of Identity operation with non-Markovian noise sources

| | | Input state | | | |
|---|---|---|---|---|---|
| | | $|00\rangle$ | $|01\rangle$ | $|10\rangle$ | $|11\rangle$ |
| Output state | $|00\rangle$ | $1-2\operatorname{Re} k(t)$ | $\operatorname{Re} k(t)$ | $\operatorname{Re} k(t)$ | 0 |
| | $|01\rangle$ | $\operatorname{Re} k(t)$ | $1-2\operatorname{Re} k(t)$ | 0 | $\operatorname{Re} k(t)$ |
| | $|10\rangle$ | $\operatorname{Re} k(t)$ | 0 | $\operatorname{Re} k(t)$ | $\operatorname{Re} k(t)$ |
| | $|11\rangle$ | 0 | $\operatorname{Re} k(t)$ | $\operatorname{Re} k(t)$ | $1-2\operatorname{Re} k(t)$ |

To examine the variability in quantum states during the execution of the Identity operation on practical NISQ devices, two platforms were used as before: 'ibm_guadalupe' provided by IBM Quantum, and IonQ accessible via Amazon Braket. The information gathered from these platforms delivers crucial understanding regarding the behavior of quantum states amidst the Idenitity operation. After performing 1,000 iterations on each platform, the probabilities of the respective quantum states have been scrupulously documented and depicted in Table 5 for 'ibm_guadalupe' and Table 6 for IonQ. These tables facilitate a side-by-side analysis of the two platforms, thereby assisting in the comprehension of how different NISQ devices can affect the fluctuations in quantum states during the Identity operation.

Table 5: Quantum state fluctuations under the Identity operation for IonQ

| | | Input state | | | |
|---|---|---|---|---|---|
| | | $|00\rangle$ | $|01\rangle$ | $|10\rangle$ | $|11\rangle$ |
| Output state | $|00\rangle$ | 0.994 | 0.015 | 0.024 | 0 |
| | $|01\rangle$ | 0.004 | 0.984 | 0 | 0.017 |
| | $|10\rangle$ | 0.002 | 0 | 0.972 | 0.007 |
| | $|11\rangle$ | 0 | 0.001 | 0.004 | 0.976 |

Table 6: Quantum state fluctuations under the Identity operation for ibm_guadalupe

| | | Input state | | | |
|---|---|---|---|---|---|
| | | $|00\rangle$ | $|01\rangle$ | $|10\rangle$ | $|11\rangle$ |
| Output state | $|00\rangle$ | 0.984 | 0.023 | 0.027 | 0 |
| | $|01\rangle$ | 0.010 | 0.972 | 0 | 0.028 |
| | $|10\rangle$ | 0.006 | 0.001 | 0.965 | 0.019 |

| | |11⟩ | 0 | 0.004 | 0.008 | 0.953 |

From the above table, the decoherence function $Re\, k(t)$ at switching time $\tau_s$ can be estimated for each NISQ machine as in the range $2.0\times 10^{-3} \leqslant Re\, k(\tau_s^{IonQ}) \leq 2.4\times 10^{-2}$ for IonQ, and $6.0\times 10^{-3} \leqslant Re\, k(\tau_s^{IBM}) \leq 2.8\times 10^{-2}$ for ibm-guadrupe, respectively.

5) **Derivation of QEM recovery operator for a SWAP operation**

The noisy SWAP evolution operator $V_{SWAP}(t)$ represented by the multiplet basis of Eq. (24) is given by

$$V_{SWAP}(t) = \begin{bmatrix} 1-2Re\,k(t) & Re\,k(t) & 0 & Re\,k(t) \\ Re\,k(t) & 1-3Re\,k(t) & Re\,k(t) & Re\,k(t) \\ 0 & Re\,k(t) & 1-2Re\,k(t) & Re\,k(t) \\ Re\,k(t) & Re\,k(t) & Re\,k(t) & 1-3Re\,k(t) \end{bmatrix}, \quad (33)$$

from Eqs. (72), (74), (76) ad (78) of the Supplementary Information.

Then the QEM recovery operator $R_{SWAP\ QEM}(t)$ is obtained from Eqs. (22) and (33) and is given by

$$R_{SWAP\ QEM}(t) = \begin{bmatrix} C & B & D & B \\ B & E & B & B \\ D & B & C & B \\ B & B & B & E \end{bmatrix}, \quad (34)$$

where

$$\begin{aligned} B &= \frac{-\alpha + 6\alpha^2 - 8\alpha^3}{1 - 10\alpha + 32\alpha^2 - 32\alpha^3}, \\ C &= \frac{1 - 8\alpha + 18\alpha^2 - 8\alpha^3}{1 - 10\alpha + 32\alpha^2 - 32\alpha^3}, \\ D &= \frac{2\alpha^2 - 8\alpha^3}{1 - 10\alpha + 32\alpha^2 - 32\alpha^3}, \\ E &= \frac{1 - 7\alpha + 14\alpha^2 - 8\alpha^3}{1 - 10\alpha + 32\alpha^2 - 32\alpha^3}, \end{aligned} \quad (35)$$

and $\alpha = Re\,k(t)$.

The QEM recovery operator can be expanded by 16 Dirac Gamma matrices $\gamma_r = I,\ \gamma_\mu, \gamma_\mu\gamma_\nu(\mu<\nu), \gamma_5\gamma_\mu, \gamma_5 = \gamma_0\gamma_1\gamma_2\gamma_3$ for $\mu = 0, 1, 2, 3$ [13,14].

Here $\gamma_i = \begin{pmatrix} 0 & \sigma_i \\ \sigma_i & 0 \end{pmatrix}$ ($i = 1, 2, 3$) and $\gamma_0 = i\begin{pmatrix} I & 0 \\ 0 & -I \end{pmatrix}$. Details of Gamma matrices are given in the Method section.

After some mathematical manipulations, we expand the QEM recovery operator as

$$R_{QEM}(t) = \frac{C+E}{2}I + D(i\gamma_1\gamma_2) + \frac{E-C}{2}(i\gamma_1\gamma_2)$$
$$+ B\left(\gamma_1 - i\gamma_2\gamma_3 - \frac{1}{2}i\gamma_5\gamma_0 - \frac{1}{2}\gamma_3\right) + \frac{D}{2}(\gamma_3 - i\gamma_5\gamma_0). \tag{36}$$

From Eq. (36), we obtain the QEM cost function as

$$c(t) = \frac{|C+E|}{2} + \frac{|C-E|}{2} + 3|B| + |D|$$
$$= \frac{|2 - 15\alpha + 32\alpha^2 - 16\alpha^3|}{2|1 - 10\alpha + 32\alpha^2 - 32\alpha^3|} + \frac{|\alpha - 4\alpha^2|}{2|1 - 10\alpha + 32\alpha^2 - 32\alpha^3|}$$
$$+ 3\frac{|-\alpha + 6\alpha^2 - 8\alpha^3|}{|1 - 10\alpha + 32\alpha^2 - 32\alpha^3|} + \frac{|2\alpha^2 - 8\alpha^3|}{|1 - 10\alpha + 32\alpha^2 - 32\alpha^3|}, \quad \alpha = Re\, k(t). \tag{37}$$

Figure 3 graphically depicts the quantum error mitigation cost function for the SWAP operation, considering the normalized gate operation time $\frac{t}{\tau_s}$ and varying coupling strengths $\Gamma_0 \omega_c \tau_s$ between a quantum system and its modeled environment of simple harmonic oscillators. The normalized gate operation time $\frac{t}{\tau_s}$ is a dimensionless parameter that compares the actual gate operation time (t) to a characteristic time scale ($\tau_s$) of the system. This normalization enables the comparison of quantum error mitigation strategies across various systems and gate operation times. The simple harmonic oscillators offer a convenient method to model these interactions and their impact on the quantum system's behavior. The cost function measures the deviation between expected and actual outcomes of quantum computations, assessing the efficacy of error reduction strategies. By normalizing gate operation time, the comparison of these strategies across different systems is possible.

Coupling strength, indicating the intensity of quantum system-environment interaction, reveals an increase in the cost function with higher strength, suggesting higher susceptibility to non-Markovian noise and environmental influences. This underscores the need for efficient error mitigation strategies for quantum systems with stronger environmental interactions. Figure 3 emphasizes the need to account for system-environment coupling strength when developing quantum error mitigation techniques, aiding in the design of tailored approaches for improved performance amidst varying environmental disturbances.

6) **Derivation of QEM recovery operator for a SWAP operation**

The noisy Identity evolution operator $V_{Id}(t)$ represented by the computational basis is given by

$$V_{Id}(t) = \begin{bmatrix} 1-2Re\ k(t) & Re\ k(t) & Re\ k(t) & 0 \\ Re\ k(t) & 1-2Re\ k(t) & 0 & Re\ k(t) \\ Re\ k(t) & 0 & 1-2Re\ k(t) & Re\ k(t) \\ 0 & Re\ k(t) & Re\ k(t) & 1-2Re\ k(t) \end{bmatrix}, \qquad (38)$$

from Eqs. (72), (74), (76) ad (78) of the Supplementary Information.

Then the QEM recovery operator $R_{Id\ QEM}(t)$ is obtained from Eqs. (22) and (33) and is given by

$$R_{Id\ QEM}(t) = \begin{bmatrix} F & H & H & G \\ H & F & G & H \\ H & G & F & H \\ G & H & H & F \end{bmatrix}, \qquad (39)$$

where

$$\begin{aligned} F &= \frac{1 - 6\alpha + 10\alpha^2 - 4\alpha^3}{1 - 8\alpha + 20\alpha^2 - 16\alpha^3}, \\ G &= \frac{2\alpha^2 - 4\alpha^3}{1 - 8\alpha + 20\alpha^2 - 16\alpha^3}, \\ H &= \frac{-\alpha + 4\alpha^2 - 4\alpha^3}{1 - 8\alpha + 20\alpha^2 - 16\alpha^3}, \end{aligned} \qquad (40)$$

and $\alpha = Re\ k(t)$.

After some mathematical manipulations, we expand the QEM recovery operator as

$$R_{QEM}(t) = FI + G(\gamma_1) + H(-i\gamma_5\gamma_0) + B(H - i\gamma_2\gamma_3). \qquad (41)$$

From Eq. (36), we obtain the QEM cost function as

$$\begin{aligned} c(t) &= |F| + |G| + 2|H| \\ &= \frac{|1 - 6\alpha + 10\alpha^2 - 4\alpha^3|}{|1 - 8\alpha + 20\alpha^2 - 16\alpha^3|} + \frac{|2\alpha^2 - 4\alpha^3|}{|1 - 8\alpha + 20\alpha^2 - 16\alpha^3|} + 2\frac{|-\alpha + 4\alpha^2 - 4\alpha^3|}{|1 - 8\alpha + 20\alpha^2 - 16\alpha^3|}, \end{aligned} \qquad (42)$$

$\alpha = Re\ k(t)$.

Figure 4 visually presents the cost function for quantum error mitigation in the Identity operation, factoring in normalized gate operation time $\frac{t}{\tau_s}$ and various coupling strengths $\Gamma_0 \omega_c \tau_s$ between a quantum system and its environment, modeled by simple harmonic oscillators. This cost function quantifies the divergence between predicted and actual quantum computation results, which helps evaluate error reduction strategies [28]. The graph suggests that stronger system-environment interactions, represented by higher coupling strengths, increase the cost function, implying a greater vulnerability to non-Markovian noise and environmental factors. Therefore, Figure 4 highlights the importance of considering system-environment coupling strength when designing effective error mitigation strategies for diverse environmental conditions.

**Discussion**

This study investigates the non-Markovian cost function for quantum error mitigation (QEM) and the representation of two-qubit operators using Dirac Gamma matrices, integral components of relativistic quantum mechanics. Our primary objective is to quantify the errors specific to individual logical gates caused by non-Markovian noise in quantum circuits, and to establish the cost function for QEM.

Our examination yielded a robust correlation between theoretical predictions and experimental data derived from a Noisy Intermediate-Scale Quantum (NISQ) device based on ion-trap technology. We discerned that the non-Markovian methodology offers a reasonably accurate depiction of the fluctuations in the output quantum states during both identity and SWAP gate operations. This evidence implies that the non-Markovian approach can be an instrumental tool for scrutinizing the impact of errors and decoherence on the output quantum states of NISQ devices. We also observed an increase in the QEM cost function as the coupling strength intensifies.

Dirac Gamma matrices constitute a collection of matrices that significantly contribute to the formulation of the relativistic quantum mechanics known as the Dirac equation. This equation outlines the behavior of spin-1/2 particles, such as electrons. The matrices showcase unique algebraic traits, encompassing anticommutativity and a distinct representation in terms of the Pauli matrices. They form a Clifford algebra and are utilized in diverse physics areas, including quantum field theory and supersymmetry.

The correlation between two-qubit gate operators and Dirac Gamma matrices stems from their mutual algebraic structure. By representing the two-qubit gate operators through Dirac Gamma matrices, we can exploit the matrices' properties to analyze and manipulate the operators more effectively. This approach could provide valuable insights into the behavior of two-qubit gates, particularly under the influence of noise and decoherence. Moreover, harnessing the algebraic properties of Dirac Gamma matrices could aid in formulating innovative error mitigation techniques and robust quantum gate designs, ultimately boosting quantum computer performance.

Our theoretical model for non-Markovian errors, premised on the Caldeira-Leggett model, aligns more accurately with the IonQ machine for the SWAP and Identity operations than the ibm_guadalupe. The likely reason is the inherent fluctuations of the qubit state in an ion trap, which align better with the Caldeira-Leggett interaction model.

To conclude, this study offers a non-Markovian approach to scrutinizing quantum state fluctuations in NISQ devices that interact with their environment, represented by simple harmonic oscillators as a source of noise. We also propose a quantitative model of a cost function for QEM using a projection operator method and advanced and retarded time propagators. This formalism can be adapted to any quantum system, contingent on the specific forms of the system, reservoir, and interaction Hamiltonians.

We have derived an analytical form of the reduced-density-operator for the output quantum states in a time-convolutionless form and juxtaposed the results with experimental data from both ion-trap and superconducting quantum computing systems. The results affirm a strong concordance between the theory and experimental outcomes. The error model and cost function, generated from the time-convolutionless equation and encompassing non-Markovian effects, could serve as bedrock elements for QEM in noisy quantum circuits.

**Methods**

**The Dirac Gamma matrices**

The $\gamma$-matrices are defined by the anti-commutation relations

$$\{\gamma_\mu, \gamma_\nu\} = 2g_{\mu\nu}, \quad \mu, \nu = 0, 1, 2, 3 \tag{40}$$

Where $g_{\mu\nu}$ is a metric tensor defined by

$$g_{11} = g_{22} = g_{33} = -g_{00} = 1, \quad g_{\mu\nu} = 0 \text{ for } \mu \neq \nu. \tag{41}$$

From the Pauli matrices $\sigma_i$, $\sigma_1 = X$, $\sigma_2 = Y$, $\sigma_3 = Z$, we construct the $4 \times 4$ Gamma matrices

$$\gamma_i = \begin{pmatrix} 0 & \sigma_i \\ \sigma_i & 0 \end{pmatrix} (i=1,2,3), \quad \gamma_0 = i\begin{pmatrix} I & 0 \\ 0 & -I \end{pmatrix}. \tag{42}$$

We introduce the 16 matrices

$$\gamma_r = I, \gamma_\mu, \gamma_\mu\gamma_\nu (\mu < \nu), \gamma_5\gamma_\mu, \gamma_5 = \gamma_0\gamma_1\gamma_2\gamma_3 \tag{43}$$

Which are subdivided into the five sets $\Gamma_\ell$ ($\ell = 1, 2, \cdots, 5$):

$$\begin{aligned}
&\Gamma_1 = I, \\
&\Gamma_2 = \gamma_\mu, (\mu = 0, \cdots, 3), \\
&\Gamma_3 = \gamma_\mu\gamma_\nu, (\mu < \nu), \\
&\Gamma_4 = \gamma_5\gamma_\mu, \\
&\Gamma_5 = \gamma_5 = \gamma_0\gamma_1\gamma_2\gamma_3.
\end{aligned} \tag{44}$$

The sets $\Gamma_1$ and $\Gamma_5$ contain one matrix each, $\Gamma_2$ and $\Gamma_4$ contain four matrices each, and $\Gamma_3$ contains six matrices. It was shown that these sixteen matrices comprising $\Gamma_\ell$ are linearly independent and therefore can represent any $4 \times 4$ matrix. Detailed expressions for 16 Gamma matrices are as follows:

$$\Gamma_1: I = \begin{pmatrix} 1 & 0 & 0 & 0 \\ 0 & 1 & 0 & 0 \\ 0 & 0 & 1 & 0 \\ 0 & 0 & 0 & 1 \end{pmatrix}, \tag{45a}$$

$\Gamma_2: \gamma_\mu \ (\mu = 0, 1, 2, 3)$

$$\gamma_0 = i\begin{pmatrix} 1 & 0 & 0 & 0 \\ 0 & 1 & 0 & 0 \\ 0 & 0 & -1 & 0 \\ 0 & 0 & 0 & -1 \end{pmatrix}, \gamma_1 = \begin{pmatrix} 0 & 0 & 0 & 1 \\ 0 & 0 & 1 & 0 \\ 0 & 1 & 0 & 0 \\ 1 & 0 & 0 & 0 \end{pmatrix}, \gamma_2 = i\begin{pmatrix} 0 & 0 & 0 & -1 \\ 0 & 0 & 1 & 0 \\ 0 & -1 & 0 & 0 \\ 1 & 0 & 0 & 0 \end{pmatrix}, \gamma_3 = \begin{pmatrix} 0 & 0 & 1 & 0 \\ 0 & 0 & 0 & -1 \\ 1 & 0 & 0 & 0 \\ 0 & -1 & 0 & 0 \end{pmatrix}, \tag{45b}$$

$\Gamma_3: \gamma_\mu \gamma_\nu \ (\mu < \nu)$

$$\gamma_0\gamma_1 = i\begin{pmatrix} 0 & 0 & 0 & 1 \\ 0 & 0 & 1 & 0 \\ 0 & -1 & 0 & 0 \\ -1 & 0 & 0 & 0 \end{pmatrix}, \ \gamma_0\gamma_2 = \begin{pmatrix} 0 & 0 & 0 & 1 \\ 0 & 0 & -1 & 0 \\ 0 & -1 & 0 & 0 \\ 1 & 0 & 0 & 0 \end{pmatrix}, \ \gamma_0\gamma_3 = i\begin{pmatrix} 0 & 0 & 1 & 0 \\ 0 & 0 & 0 & -1 \\ -1 & 0 & 0 & 0 \\ 0 & 1 & 0 & 0 \end{pmatrix}, \quad (45c)$$

$$\gamma_1\gamma_2 = i\begin{pmatrix} 1 & 0 & 0 & 0 \\ 0 & -1 & 0 & 0 \\ 0 & 0 & 1 & 0 \\ 0 & 0 & 0 & -1 \end{pmatrix}, \ \gamma_1\gamma_3 = \begin{pmatrix} 0 & -1 & 0 & 0 \\ 1 & 0 & 0 & 0 \\ 0 & 0 & 0 & -1 \\ 0 & 0 & 1 & 0 \end{pmatrix}, \ \gamma_2\gamma_3 = i\begin{pmatrix} 0 & 1 & 0 & 0 \\ 1 & 0 & 0 & 0 \\ 0 & 0 & 0 & 1 \\ 0 & 0 & 1 & 0 \end{pmatrix},$$

$\Gamma_4: \gamma_5 \gamma_\mu \ (\mu = 0,1,2,3)$

$$\gamma_5\gamma_0 = i\begin{pmatrix} 0 & 0 & 1 & 0 \\ 0 & 0 & 0 & 1 \\ 1 & 0 & 0 & 0 \\ 0 & 0 & 0 & 0 \end{pmatrix}, \ \gamma_5\gamma_1 = \begin{pmatrix} 0 & -1 & 0 & 0 \\ -1 & 0 & 0 & 0 \\ 0 & 0 & 0 & 1 \\ 0 & 0 & 1 & 0 \end{pmatrix}, \ \gamma_5\gamma_2 = i\begin{pmatrix} 0 & 1 & 0 & 0 \\ -1 & 0 & 0 & 0 \\ 0 & 0 & 0 & 1 \\ 0 & 0 & 1 & 0 \end{pmatrix}, \ \gamma_5\gamma_3 = \begin{pmatrix} -1 & 0 & 0 & 0 \\ 0 & 1 & 0 & 0 \\ 0 & 0 & 1 & 0 \\ 0 & 0 & 0 & -1 \end{pmatrix}, \quad (45d)$$

$\Gamma_5: \gamma_5 =$
$$\gamma_5 = \begin{pmatrix} 0 & 0 & -1 & 0 \\ 0 & 0 & 0 & -1 \\ 1 & 0 & 0 & 0 \\ 0 & 1 & 0 & 0 \end{pmatrix}. \quad (45e)$$

33. Zou, J., Bosco, S., and Loss, D. Spatially correlated classical and quantum noise in driven qubits: The good, the bad, and the ugly. arXiv:2308.03054 (2023).

## Acknowledgments

The author thanks Mr. Byeongyong Park for his help with the experiment. This work was supported by Korea National Research Foundation (NRF) grant No. NRF-2023R1A2C1003570, ICT R&D program of MSIT/IITP 2021-0-01810, RS-2023-00225385, AFOSR grant FA2386-21-1-0089, AFOSR grant FA2386-22-1-4052, and Amazon Web Services. We acknowledge the use of IBM Quantum services for this work. The views expressed are those of the author, and do not reflect the official policy or position of IBM or the IBM Quantum team. This work was also supported by the National Quantum Laboratory at Maryland (QLab).

## Data availability

The data generated during the current study are available from the corresponding author on reasonable request.

## Competing interests

The authors declare no competing interests.

**Figure legends**

**Figure 1.** The open quantum system is composed of a two-state quantum subsystem $H_S$, an environment $H_B$, and their interaction. The interaction $H_{int}$ between the quantum subsystem and the environment plays a crucial role in determining the system's dynamics and the noise sources affecting its performance. This comprehensive view of the open quantum system provides a framework for understanding and analyzing error sources and devising effective error mitigation strategies.

**Figure 2.** Plot of $Re\, k(t)$ vs $\frac{t}{\tau_s}$, where $\tau_s$ is the switching time for the parameters $\Gamma_0 \omega_c \tau_s$ in the range $7.0 \times 10^{-4} \leqslant \Gamma_0 \omega_c \tau_s \leq 7.0 \times 10^{-3}$.

**Figure 3**. A graphical representation of the quantum error mitigation cost function for SWAP operation with respect to the normalized gate operation time $\frac{t}{\tau_s}$ for varying coupling strengths $\Gamma_0 \omega_c \tau_s$ between a quantum system and its environment.

**Figure 4**. A graphical representation of the quantum error mitigation cost function for Identity operation with respect to the normalized gate operation time $\frac{t}{\tau_s}$ for varying coupling strengths $\Gamma_0 \omega_c \tau_s$ between a quantum system and its environment.

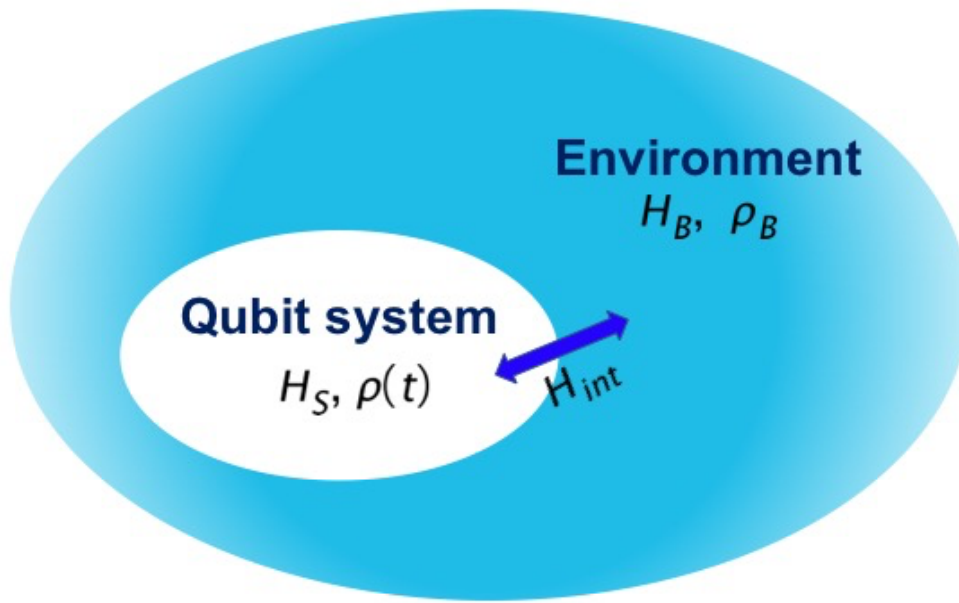

**Fig. 1**

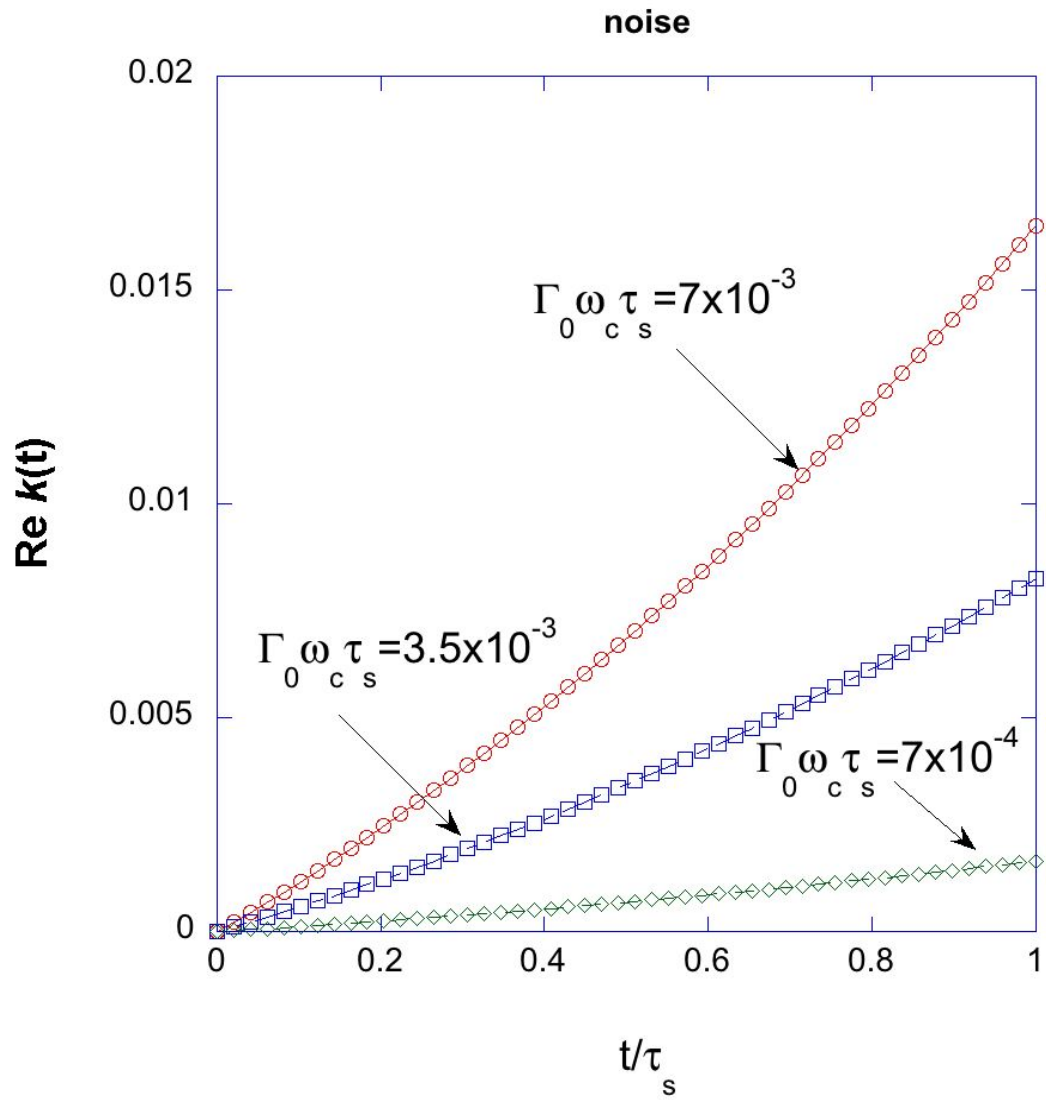

**Fig. 2**

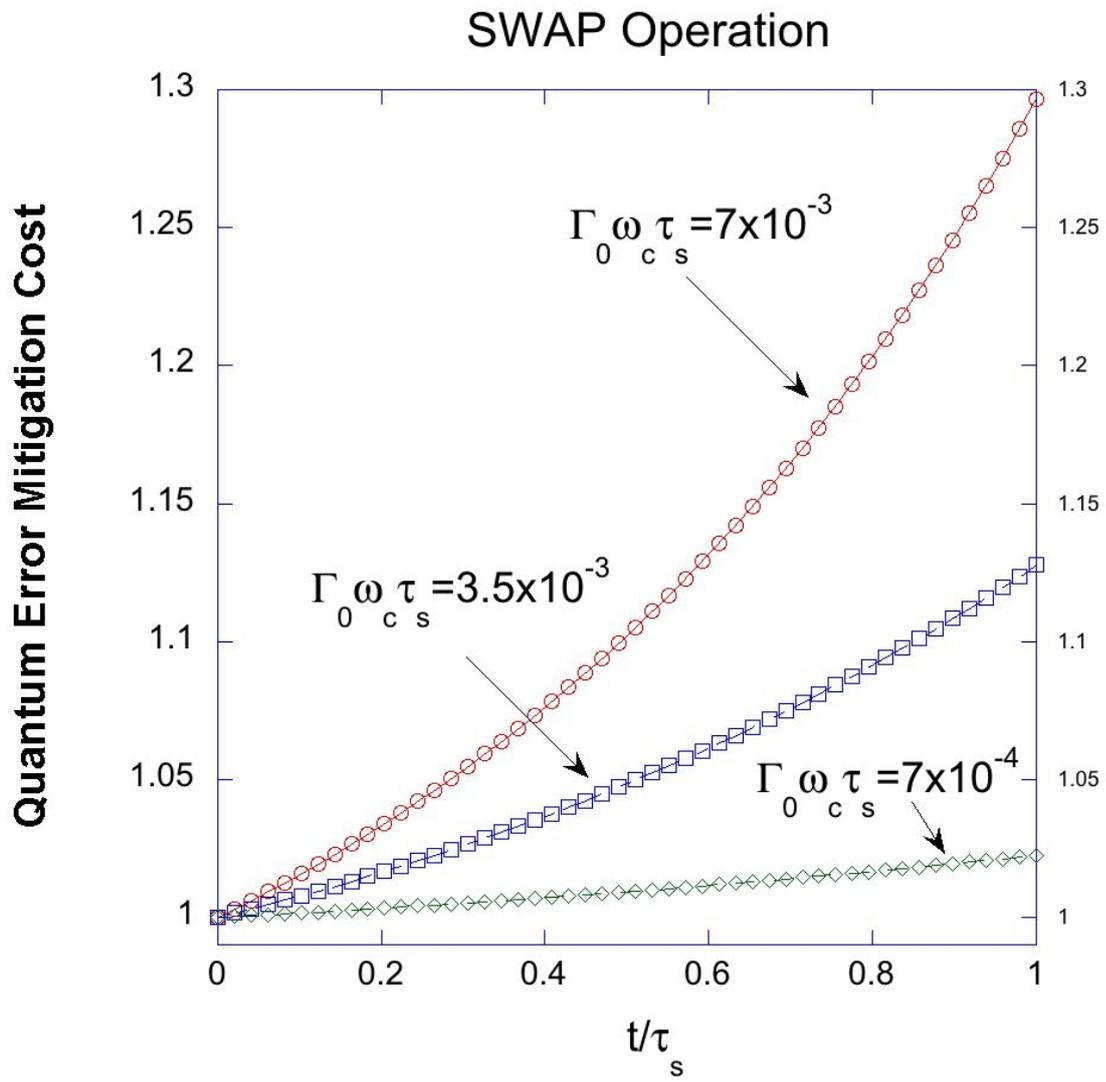

**Fig. 3**

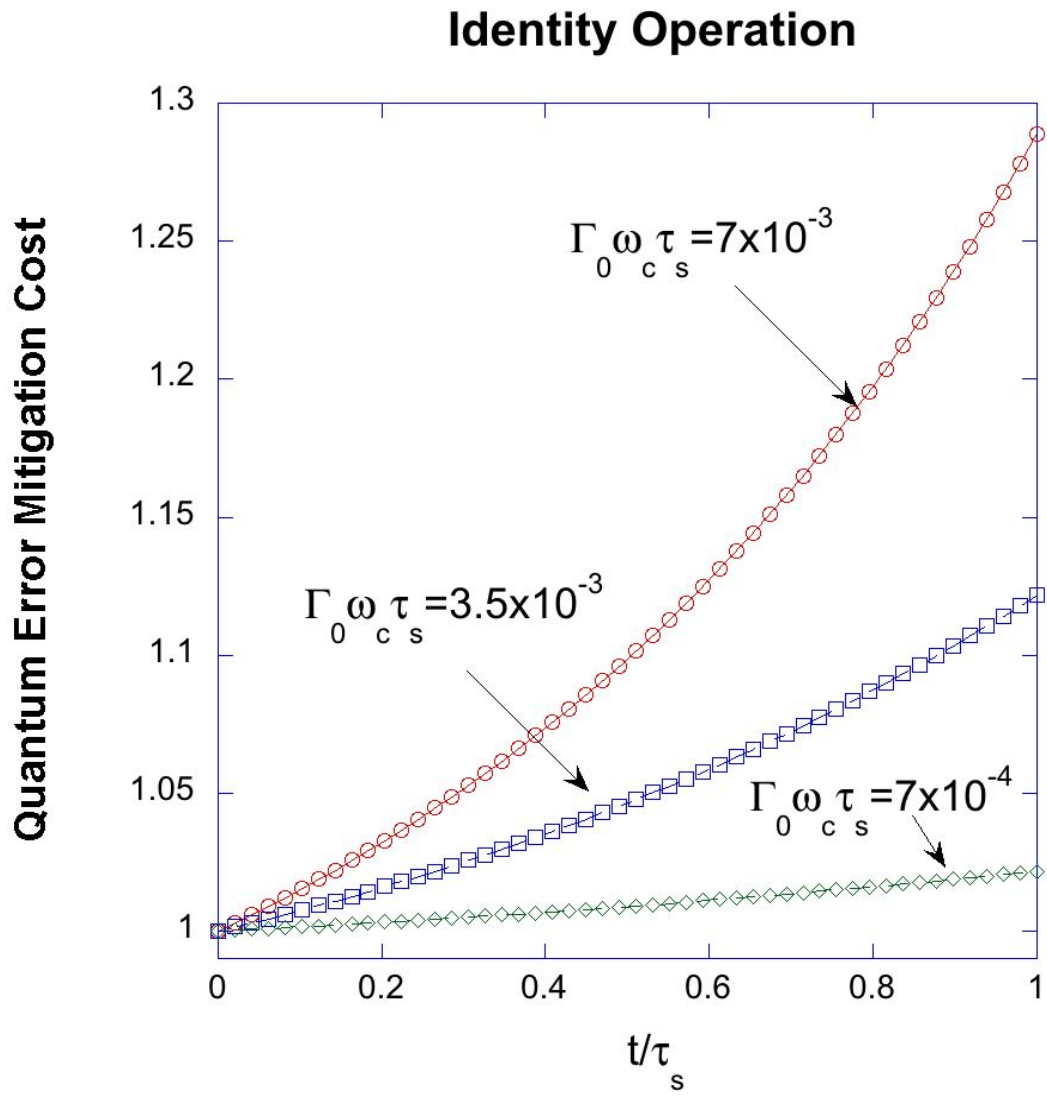

**Fig. 4**

# Supplementary Information for
# Non-Markovian cost function for quantum error mitigation with Dirac Gamma matrices representation

Doyeol Ahn[1,*]

[1]Department of Electrical and Computer Engineering,
University of Seoul, 163 Seoulsiripdae-ro, Tongdaimoon-gu, Seoul 02504, Republic of Korea

*Corresponding author: dahn@uos.ac.kr

## I. Time-convolutionless evolution of a non-Markovian reduced-density operator

The total Hamiltonian for an open two-state system is given by [1-5]

$$\widehat{H}_T = \widehat{H}_S(t) + \widehat{H}_B + \widehat{H}_{int} \tag{1}$$

where $\widehat{H}_S(t)$ is the system Hamiltonian for a two-state system, $\widehat{H}_B$ the Hamiltonian acting on the reservoir or an environment, $\widehat{H}_{int}$ is the interaction between the system and the environment (as depicted in figure 1).

The equation of motion for the total density operator $\rho_T(t)$ of the total system is given by a quantum Liouville equation [1]

$$\frac{d}{dt}\rho_T(t) = -i[\widehat{H}_T(t), \rho_T(t)] = -i\widehat{L}_T(t)\rho_T(t), \tag{2}$$

where

$$\widehat{L}_T(t) = \widehat{L}_S(t) + \widehat{L}_B(t) + \widehat{L}_{int}(t) \tag{3}$$

is the Liouville super operator in one-to-one correspondence with the Hamiltonian. Here, we use the unit in which $\hbar = 1$. In order to derive an equation and to solve for a system alone, it is convenient to use the projection operators which decompose the total system by eliminating the degrees of freedom for the reservoir. We define thine-independent projection operator $\underline{P}$ and $\underline{Q}$ given by [1,2]

$$\underline{P}X = \rho_B tr_B X, \underline{Q} = 1 - \underline{P} \tag{4}$$

for any dynamical variable $X$. Here $tr_B$ denotes a partial trace over the quantum reservoir and $\rho_B$ is the density matrix of the reservoir. The projection operators satisfy the following operator identities:

$$\underline{P}^2 = \underline{P}, \ \underline{Q}^2 = \underline{Q} \ \underline{P}\,\underline{Q} = \underline{Q}\underline{P} = 0, \tag{5a}$$

$$\underline{P}\widehat{L}_S(t) = \widehat{L}_S(t)\underline{P}, \ \underline{P}\widehat{L}_B = \widehat{L}_B\underline{P} = 0, \ \underline{P}\widehat{L}_{int}\underline{P} = 0, \tag{5b}$$

and $\quad \underline{Q}\widehat{L}_S(t) = \widehat{L}_S(t)\underline{Q}, \ \underline{Q}\widehat{L}_B = \widehat{L}_B\underline{Q} = \widehat{L}_B. \tag{5c}$

The information of the system is then contained in the reduced density operator

$$\rho(t) = tr_B \rho_T(t) = tr_B \underline{P} \rho_T(t). \tag{6}$$

If we multiply equation (2) by $\underline{P}$ and $\underline{Q}$ on the left side, we obtain coupled equations for $\underline{P}\rho_T(t)$ and $\underline{Q}\rho_T(t)$ as follows:

$$\frac{d}{dt}\underline{P}\rho_T(t) = -i\underline{P}\widehat{L}_T(t)\underline{P}\rho_T(t) - -i\underline{P}\widehat{L}_T(t)\underline{Q}\rho_T(t), \tag{7a}$$

and

$$\frac{d}{dt}\underline{Q}\rho_T(t) = -i\underline{Q}\widehat{L}_T(t)\underline{Q}\rho_T(t) - -i\underline{Q}\widehat{L}_T(t)\underline{P}\rho_T(t), \tag{7b}$$

where we have modified the Eq. (2) as

$$\frac{d}{dt}\rho_T(t) = -i\widehat{L}_T(t)\rho_T(t) = --i\widehat{L}_T(t)\left(\underline{P}+\underline{Q}\right)\rho_T(t).$$

We assume that the system was turned on at $t=0$ and the input state prepared at $t=0$ was isolated with the reservoir such that $\underline{Q}\rho_T(0)=0$. The formal solution of (7b) is given by

$$\begin{aligned}
\underline{Q}\rho_T(t) &= -i\int_0^t \underline{H}(t,\tau)\,\underline{Q}\widehat{L}_T(\tau)\,\underline{P}\rho_T(\tau)\,d\tau + \underline{H}(t,0)\,\underline{Q}\rho_T(0) \\
&= -i\int_0^t \underline{H}(t,\tau)\,\underline{Q}\widehat{L}_T(\tau)\,\underline{P}\rho_T(\tau)\,d\tau.
\end{aligned} \tag{8}$$

where the projected propagator $\underline{H}(t,\tau)$ of the total system is defined by

$$\underline{H}(t,\tau) = \underline{T}exp\left[-i\int_\tau^t \underline{Q}\widehat{L}_T(s)\,\underline{Q}ds\right]. \tag{9}$$

Here $\underline{T}$ is the time-ordering operator. We also introduce an anti-time evolution operator $\underline{G}(t,\tau)$ which is defined by

$$\underline{G}(t,\tau) = \underline{T}^C \exp\left[i\int_\tau^t \widehat{L}_T(s)\,ds\right], \tag{10}$$

such that $\rho_T(\tau) = \underline{G}(t,\tau)\rho_T(t)$. \hfill (11)

It is not very difficult to show [1] that

$$\underline{G}(t,\tau)\,\underline{G}(s,t) = \underline{G}(s,\tau),\ \underline{H}(t,\tau)\,\underline{H}(\tau,s) = \underline{H}(t,s). \tag{12}$$

Here $\underline{T}^C$ is anti-time-ordering operator.

From Eqs. (8)-(11), we obtain

$$\underline{Q}\rho_T(t) = -i\int_0^t \underline{H}(t,\tau)\,\underline{Q}\widehat{L}_T(\tau)\,\underline{P}\,\underline{G}(t,\tau)\left(\underline{P}+\underline{Q}\right)\rho_T(t) \tag{13}$$

which is obviously in time-convolutionless form. Eq. (13) can be rewritten as

$$\left[1 + i\int_0^t \underline{H}(t,\tau)\, \underline{Q}\, \widehat{L}_T(\tau)\, \underline{P}\, \underline{G}(t,\tau)\, d\tau\right] \underline{Q}\rho_T(t)$$
$$= -i\int_0^t d\tau \underline{H}(t,\tau)\, \underline{Q}\, \widehat{L}_T(\tau)\, \underline{P}\, \underline{G}(t,\tau)\, \underline{P}\, \rho_T(t). \tag{14}$$

We introduce new super operators $\hat{g}(t)$ and $\hat{\theta}(t)$ which are given by

$$\hat{g}(t) = 1 + i\int_0^t d\tau \underline{H}(t,\tau)\, \underline{Q}\widehat{L}_T(\tau)\, \underline{P}\, \underline{G}(t,\tau) = \hat{\theta}^{-1}(t). \tag{15}$$

Then from Eqs. (13)-(15), we obtain

$$\underline{Q}\rho_T(t) = \{\hat{\theta}(t) - 1\}\underline{P}\rho_T(t) \tag{16}$$

and

$$\frac{d}{dt}\underline{P}\rho_T(t) + i\underline{P}\widehat{L}_T(t)\, \underline{P}\, \rho_T(t) = -i\underline{P}\widehat{L}_T(t)\, \{\hat{\theta}(t) - 1\}\underline{P}\, \rho_T(t). \tag{17}$$

After some mathematical manipulations, we obtain the formal solution of Eq. (17), which is given by

$$\underline{P}\rho_T(t) = \widehat{U}(t,0)\,\underline{P}\rho_T(0) - i\int_0^t ds\, \widehat{U}(t,s)\, \underline{P}\widehat{L}_T(s)\,\{\hat{\theta}(s) - 1\}\underline{P}\rho_T(s)$$

$$= \widehat{U}(t,0)\,\underline{P}\rho_T(0) - i\int_0^t ds\, \widehat{U}(t,s)\, \underline{P}\widehat{L}_T(s)\,\{\hat{\theta}(s) - 1\}\underline{P}\, \underline{G}(t,s)\rho_T(t)$$

$$= \widehat{U}(t,0)\,\underline{P}\rho_T(0) - i\int_0^t ds\, \widehat{U}(t,s)\, \underline{P}\widehat{L}_T(s)\,\{\hat{\theta}(s) - 1\}\underline{P}\, \underline{G}(t,s)\underline{P}\rho_T(t)$$

$$\qquad - i\int_0^t ds\, \widehat{U}(t,s)\, \underline{P}\widehat{L}_T(s)\,\{\hat{\theta}(s) - 1\}\underline{P}\, \underline{G}(t,s)\underline{Q}\rho_T(t) \tag{18}$$

$$= \widehat{U}(t,0)\,\underline{P}\rho_T(0) - i\int_0^t ds\, \widehat{U}(t,s)\, \underline{P}\widehat{L}_T(s)\,\{\hat{\theta}(s) - 1\}\underline{P}\, \underline{G}(t,s)\underline{P}\rho_T(t)$$

$$\qquad - i\int_0^t ds\, \widehat{U}(t,s)\, \underline{P}\widehat{L}_T(s)\,\{\hat{\theta}(s) - 1\}\underline{P}\, \underline{G}(t,s)\{\hat{\theta}(t) - 1\}\underline{P}\rho_T(t)$$

$$= \widehat{U}(t,0)\,\underline{P}\rho_T(0) - i\int_0^t ds\, \widehat{U}(t,s)\, \underline{P}\widehat{L}_T(s)\,\{\hat{\theta}(s) - 1\}\underline{P}\, \underline{G}(t,s)\hat{\theta}(t)\underline{P}\rho_T(t),$$

where

$$\widehat{U}(t,\tau) = \underline{T}exp\left[-i\int_\tau^t ds\, \underline{P}\widehat{L}_T(s)\, \underline{P}\right] \tag{19}$$

is the projected propagator of the total system.

From, $\underline{P}\widehat{L}_T(t)\underline{P} = \underline{P}\left(\widehat{L}_S(t) + \widehat{L}_B + \widehat{L}_{int}\right)\underline{P} = \underline{P}\,\widehat{L}_S(t)\,\underline{P} = \widehat{L}_S(t)\,\underline{P}^2 = \widehat{L}_S(t)\,\underline{P}$, we obtain

$$\widehat{U}(t,s)\,\underline{P} = \underline{T}exp\left[-i\int_s^t d\tau \underline{P}\widehat{L}_T(\tau)\,\underline{P}\right]\underline{P}$$

$$= \underline{T}\,exp\left[-i\int_s^t d\tau \widehat{L}_S(\tau)\,\underline{P}\right]\underline{P} \tag{20}$$

$$= \underline{T}\,exp\left[-i\int_s^t d\tau \widehat{L}_S(\tau)\right]\underline{P} = \widehat{U}_S(t,s)\,\underline{P}.$$

Here $\widehat{U}_S(t,s)$ is the propagator of the system.

From Eqs. (4) and (6), we get

$$\widehat{U}(t,s)\underline{P}\widehat{L}_T(s)\{\widehat{\theta}(s)-1\}\underline{P}\,\underline{G}(t,s)\widehat{\theta}(t)\,\underline{P}\rho_T(t)$$
$$=\widehat{U}_S(t,s)\rho_B\,tr_B\{\underline{P}\widehat{L}_T(s)\{\widehat{\theta}(s)-1\}\rho_B\}tr_B\{\underline{G}(t,s)\widehat{\theta}(t)\rho_B\}\rho(t)$$

By the way, by substituting Eq. (7) into Eqs. (18)-(20), we obtain

$$\rho_B\rho(t)=\rho_B\widehat{U}_S(t,0)\rho(0)-i\rho_B\int_0^t ds\widehat{U}_S(t,s)tr_B[\widehat{L}_T(s)\{\widehat{\theta}(s)-1\}\rho_B]tr_B[\underline{G}(t,s)\widehat{\theta}(t)\rho_B]\rho(t),$$

(21)

or

$$\left(1+i\int_0^t ds\widehat{U}_S(t,s)tr_B[\widehat{L}_T(s)\{\widehat{\theta}(s)-1\}\rho_B]tr_B[\underline{G}(t,s)\widehat{\theta}(t)\rho_B]\right)\rho(t)$$
$$=\widehat{U}_S(t,s)\rho(t).$$

(22)

If we define $\widehat{W}(t)$ by

$$\widehat{W}(t)=1+i\int_0^t ds\widehat{U}_S(t,s)tr_B[\widehat{L}_T(s)\{\widehat{\theta}(s)-1\}\rho_B]tr_B[\underline{G}(t,s)\widehat{\theta}(t)\rho_B],\qquad(23)$$

Then, the evolution operator for the reduced density operator $\rho(t)$ is given by

$$\rho(t)=\widehat{W}^{-1}(t)\widehat{U}_S(t,0)\rho(0)=\widehat{V}(t)\rho(0),\qquad(24)$$

where the super-operator $\widehat{V}(t)$ for the evolution of the reduced density operator is defined by $\widehat{V}(t)=\widehat{W}^{-1}(t)\widehat{U}_S(t,0)$. Within the Born approximation, we have

$$\widehat{V}^{(2)}(t)=\left\{1-\int_0^t ds\int_0^s d\tau\widehat{U}_S(t,s)tr_B[\widehat{L}_{int}\widehat{U}_o(s,\tau)\widehat{L}_{int}\widehat{U}_o^{-1}(s,\tau)\rho_B]\widehat{U}_o^{-1}(t,s)\right\}\widehat{U}_S(t,0),\qquad(25)$$

where

$$\widehat{U}_o(t)=\underline{T}exp\left[-i\int_0^t d\tau\left(\widehat{L}_S(\tau)+\widehat{L}_B\right)\right]$$
$$=exp(-it\widehat{L}_B)\underline{T}exp\left[-i\int_0^t d\tau\left(\widehat{L}_S(\tau)\right)\right]\qquad(26)$$
$$=\widehat{U}_B(t)\widehat{U}_S(t).$$

A detailed derivation of Eq. (25) is given in the below:

We start with Eq. (23)

$$\widehat{W}(t)=1+i\int_0^t ds\widehat{U}_S(t,s)tr_B[\widehat{L}_T(s)\{\widehat{\theta}(s)-1\}\rho_B]tr_B[\underline{G}(t,s)\widehat{\theta}(t)\rho_B].\qquad(23)$$

We define

$$\widehat{\Sigma}(t)=1-\widehat{\theta}(t)$$
$$=-i\int_0^t d\tau\underline{H}(t,\tau)\underline{Q}\,\widehat{L}_T(\tau)\underline{P}\,\underline{G}(t,\tau),\qquad(27)$$

then
$$\widehat{\theta}^{-1}(t) = 1 - \widehat{\Sigma}(t),$$
$$\widehat{\theta}(t) - 1 = \frac{\widehat{\Sigma}(t)}{1 - \widehat{\Sigma}(t)}.$$
(28)

Also, we have
$$\underline{P}\,\widehat{L}_T(t)\{\widehat{\theta}(t) - 1\}$$
$$= \underline{P}\left(\widehat{L}_S(t) + \widehat{L}_B + \widehat{L}_{int}\right)\widehat{\Sigma}(t)\{1 - \widehat{\Sigma}(t)\}^{-1}.$$

The detailed expression for $\widehat{\Sigma}(t)$ becomes
$$\widehat{\Sigma}(t) = -i\int_0^t d\tau \underline{H}(t,\tau)\,\underline{Q}\,\widehat{L}_T(\tau)\,\underline{P}\,\underline{G}(t,\tau)$$
$$= -i\int_0^t d\tau \underline{H}(t,\tau)\,\underline{Q}\,\widehat{L}_{int}(\tau)\,\underline{P}\,\underline{G}(t,\tau) \qquad (29)$$
$$= -i\int_0^t d\tau \widehat{U}_o(t)\,\underline{S}(t,\tau)\,\widehat{U}_o^{-1}(\tau)\,\underline{Q}\,\widehat{L}_T(\tau)\,\underline{P}\,\widehat{U}_o(\tau)\,\underline{R}(t,\tau)\,\widehat{U}_o^{-1}(t),$$

where
$$\underline{R}(t,\tau) = \underline{T}^C \exp\left[i\int_\tau^t ds\,\widehat{U}_o^{-1}(s)\,\widehat{L}_{int}\,\widehat{U}_o(s)\right] \qquad (30)$$

is the evolution operator in the interaction picture and
$$\underline{S}(t,\tau) = \underline{T}\exp\left[i\int_\tau^t ds\,\underline{Q}\,\widehat{U}_o^{-1}(s)\,\widehat{L}_{int}\,\widehat{U}_o(s)\,\underline{Q}\right] \qquad (31)$$

is the projected propagator of the total system in the interaction picture. Then Eq. (23) can be rewritten as
$$\widehat{W}(t) = 1 + i\int_0^t ds\,\widehat{U}_S(t,s)\,tr_B\!\left[\widehat{L}_{int}\widehat{\Sigma}(s)\{1-\widehat{\Sigma}(s)\}^{-1}\rho_B\right]tr_B\!\left[\widehat{U}_o(s)\underline{R}(t,s)\widehat{U}_o^{-1}\{1-\widehat{\Sigma}(t)\}^{-1}\rho_B\right].$$
(32)

where
$$\underline{R}(t,\tau) = \underline{T}^C \exp\left[i\int_\tau^t ds\,\widehat{U}_o^{-1}(s)\,\widehat{L}_{int}\,\widehat{U}_o(s)\right] \qquad (33)$$

is the evolution operator in the interaction picture and
$$\underline{S}(t,\tau) = \underline{T}\exp\left[i\int_\tau^t ds\,\underline{Q}\,\widehat{U}_o^{-1}(s)\,\widehat{L}_{int}\,\widehat{U}_o(s)\,\underline{Q}\right] \qquad (34)$$

is the projected propagator of the total system in the interaction picture. Then Eq. (23) can be rewritten as
$$\widehat{W}(t) = 1 + i\int_0^t ds\,\widehat{U}_S(t,s)\,tr_B\!\left[\widehat{L}_{int}\widehat{\Sigma}(s)\{1-\widehat{\Sigma}(s)\}^{-1}\rho_B\right]tr_B\!\left[\widehat{U}_o(s)\underline{R}(t,s)\widehat{U}_o^{-1}\{1-\widehat{\Sigma}(t)\}^{-1}\rho_B\right].$$

$$\tag{35}$$

Born approximation of Eq. (35) leads to

$$\widehat{W}^{(2)}(t) = 1 + i\int_0^t ds\, \widehat{U}_S(t,s)\, tr_B\!\left[\widehat{L}_{int}\widehat{\Sigma}^{(1)}(s)\rho_B\right] tr_B\!\left[\widehat{U}_o(s)\widehat{U}_o^{-1}\rho_B\right], \tag{36}$$

where

$$\begin{aligned}
\widehat{\Sigma}^{(1)}(s) &= -i\int_0^s ds\, \widehat{U}_o(s)\widehat{U}_o^{-1}(\tau)\underline{Q}\widehat{L}_{int}\underline{P}\,\widehat{U}_o(\tau)\widehat{U}_o^{-1}(s) \\
&= -i\int_0^s ds\, \widehat{U}_o(s)\widehat{U}_o^{-1}(\tau)\widehat{L}_{int}\widehat{U}_o(\tau)\widehat{U}_o^{-1}(s) \\
&= -i\int_0^s ds\, \widehat{U}_o(s,\tau)\widehat{L}_{int}\widehat{U}_o^{-1}(s,\tau),
\end{aligned} \tag{37}$$

using the ansatz $\underline{P}\widehat{L}_{int}\underline{P}=0$. From Eqs. (36) and (37), we get

$$\begin{aligned}
\widehat{W}^{-1(2)}(t) &= 1 - i\int_0^t ds\, \widehat{U}_S(t,s)\, tr_B\!\left[\widehat{L}_{int}\widehat{\Sigma}^{(1)}(s)\rho_B\right] tr_B\!\left[\widehat{U}_o(s)\widehat{U}_o^{-1}\rho_B\right] \\
&= 1 - i\int_0^t ds\, \widehat{U}_S(t,s)\, tr_B\!\left[\widehat{L}_{int}\widehat{\Sigma}^{(1)}(s)\rho_B\right]\widehat{U}_S^{-1}(t,s) \\
&= 1 - \int_0^t ds\int_0^s d\tau\, \widehat{U}_S(t,s)\, tr_B\!\left[\widehat{L}_{int}\widehat{U}_o(s,\tau)\widehat{L}_{int}\widehat{U}_o^{-1}(s,\tau)\rho_B\right]\widehat{U}_S^{-1}(t,s).
\end{aligned} \tag{38}$$

By substituting Eq. (38) into $\widehat{V}^{(2)}(t)=\widehat{W}^{-1(2)}(t)\widehat{U}_S(t,0)$, we obtain Eq. (25). After some mathematical manipulations, $\widehat{V}^{(2)}(t)\rho(0)$ becomes

$$\begin{aligned}
\rho(t) &= \widehat{V}^{(2)}(t)\rho(0) \\
&= \widehat{U}_S(t)\rho(0) - \int_0^t ds\int_0^s d\tau\, d\tau\, tr_B\!\left[\widehat{H}_{int}\widehat{H}_{int}(\tau-s)\rho_B\rho(-s)\right] + \int_0^t ds\int_0^s d\tau\, tr_B\!\left[\widehat{H}_{int}(\tau-s)\rho_B\rho(-s)\rho_B\widehat{H}_{int}\right] \\
&\quad + \int_0^t ds\int_0^s d\tau\, tr_B\!\left[\widehat{H}_{int}\rho_B\rho(-s)\widehat{H}_{int}(\tau-s)\right] - \int_0^t ds\int_0^s d\tau\, tr_B\!\left[\rho_B\rho(-s)\widehat{H}_{int}(\tau-s)\widehat{H}_{int}\right].
\end{aligned} \tag{39}$$

Here $\widehat{H}_{int}(\tau-s)$ and $\rho(-s)$ are Heisenberg operators defined by

$$\widehat{H}_{int}(\tau-s) = \exp\!\left(i\widehat{H}_o(\tau-s)\right)\widehat{H}_{int}\exp\!\left(-i\widehat{H}_o(\tau-s)\right),$$

$$\rho(-s) = \exp\!\left(-i\int_0^s dt\,\widehat{H}_S(t)\right)\rho(0)\exp\!\left(i\int_0^s dt\,\widehat{H}_S(t)\right),$$

respectively.

**Non-Markovian evolution of two-qubit gate operation**

In this work, we focus on the two-qubit gate operations and model the interaction of the quantum system with the environment during the gate operation by a Caldeira-Leggett model [3-5] where a set of harmonic oscillators are coupled linearly with the system spin by

$$\widehat{H}_{int} = \lambda \sum_{i=1,2,\, j=1,2,3} \vec{S}_i \cdot \vec{b}_i,\ S_i^j = \frac{\hbar}{2} \sigma^j \tag{40}$$

where $\sigma_i$ is the Pauli matrices $\sigma_1 = X$, $\sigma_2 = Y$, $\sigma_3 = Z$ and $b_i^j$ is the fluctuating quantum field associated the i$^{th}$ qubit, whose motion is governed by the harmonic-oscillator Hamiltonian.

In the evaluation of Eq. (39), we obtain the following relations [1]:

$$\begin{aligned} tr_B\{b_k^l(t)\, b_i^j \rho_B\} &= \delta_{ik} \delta_{jl} [\Gamma(t) + i\Delta(t)], \\ tr_B\{b_i^j b_k^l(t)\, \rho_B\} &= \delta_{ik} \delta_{jl} [\Gamma(t) - i\Delta(t)], \end{aligned} \tag{41}$$

where

$$\Gamma(t) + i\Delta(t) = \frac{\lambda^2}{\pi} \int_0^\infty J(\omega) \left\{ \exp(-i\omega t) + \coth\left(\frac{\omega}{2k_B T}\right) \cos \omega t \right\}. \tag{42}$$

Here $J(\omega)$ is the ohmic damping given by $J(\omega) = \theta(\omega_c - \omega)\eta\omega$, $\Gamma$ is the decoherence rate of the qubit system.

We evaluate the reduced-density-operator in the multiplet basis representation [1]

$$\rho(t) = \sum_{a,b} \rho_{ab}(t)\, e_{ab},\ e_{ab} = |a\rangle\langle b|,\ a,b = 1,2,3,4, \tag{43}$$

where $e_{ab}$ is the multiplet states. The inner product between the multiplet basis is defined by

$$(e_{ab},\, e_{cd}) = tr\left[e_{ab}^\dagger e_{cd}\right] = \delta_{ac} \delta_{bd}. \tag{44}$$

Then, from Eqs. (39)-(44), we obtain the matrix component of the reduced-density-operator as

$$\begin{aligned} \rho_{ab}(t) &= V_{ab|cd}(t)\, \rho(0), \\ V_{ab|cd}(t) &= \exp\left[-it(E_a - E_b)\right] \left\{ \delta_{ac}\delta_{bd} - \left[\delta_{bd} \sum_{a'} M_{aa'a'c} - M_{acdb}\right] k(t) - \left[\delta_{ac} \sum_{a'} M_{aa'a'b} - M_{acdb}\right] k^*(t) \right\} \end{aligned} \tag{45}$$

where

$$\begin{aligned} M_{abcd} &= \sum_{i,j} \langle a|S_i^j|b\rangle \langle c|S_i^j|d\rangle \\ &= \frac{1}{4} \sum_{i=1,2} \left\{ \langle a|X_i|b\rangle \langle c|X_i|d\rangle + \langle a|Y_i|b\rangle \langle c|Y_i|d\rangle + \langle a|Z_i|b\rangle \langle c|Z_i|d\rangle \right\}, \end{aligned} \tag{46}$$

and
$$k(t) = \frac{2}{\pi}\Gamma_o\left\{\frac{\pi}{2}\omega_c t + \int_0^{\frac{t}{\tau_s}} Si(\omega_c \tau_s t)\, dt\right\} \quad (47)$$

$$+ i\Delta_o\left\{\frac{t}{\tau_s} - \frac{1}{\omega_c \tau_s}\left(\frac{\pi}{2}\omega_c \tau_s + Si(\omega_c t)\right)\right\}.$$

Here $\omega_c$ is the high frequency cutoff, $\tau$ is the switching time, $\Gamma_0 = \lambda^2 \eta k_B T \tau_s$ and
$\Delta_o = \dfrac{\lambda^2 \eta \omega_c \tau_s}{\pi}$.

We now study the non-Markovian errors associated with two-qubit gate operations. Here, $Si(x) = \int_0^x dt \dfrac{\sin t}{t}$, is a sine integral. For reference, we also note the cosine integral, $Ci(x) = -\int_x^\infty dt \dfrac{\cos t}{t}$.

Derivation of Equations (45) to (47) is given in the below:

We define the operator integrand $\widehat{B}$ as follows:

$$\begin{aligned}\widehat{B} &= \widehat{L}_{int}\widehat{U}_o(s,\tau)\widehat{L}_{int}\widehat{U}_o^{-1}(s,\tau)\rho_B\widehat{U}_S^{-1}(s,t)\widehat{U}_S(t)\rho(0)\\ &= \widehat{L}_{int}\widehat{U}_o(s,\tau)\widehat{L}_{int}\widehat{U}_o^{-1}(s,\tau)\rho_B\widehat{U}_S(s)\rho(0)\\ &= \widehat{L}_{int}\widehat{U}_S(s,\tau)\widehat{U}_B(s,\tau)\widehat{L}_{int}\widehat{U}_B(\tau,s)\widehat{U}_S(\tau,s)\rho_B\widehat{U}_S(s)\rho(0)\\ &= \widehat{L}_{int}\widehat{U}_S(s,\tau)\widehat{U}_B(s,\tau)\widehat{L}_{int}\rho_B(s-\tau)\rho(-\tau).\end{aligned} \quad (48)$$

Here, we used the relation

$$\begin{aligned}&\widehat{U}_o^{-1}(s,\tau)\rho_B\widehat{U}_S(s)\rho(0)\\ &= \widehat{U}_o(\tau,s)\rho_B\widehat{U}_S(s)\rho(0)\\ &= exp[-i\widehat{L}_o(\tau-s)]\rho_B exp[-is\widehat{L}_S]\rho(0)\\ &= exp[-i\widehat{H}_o(\tau-s)]\rho_B exp[-is\widehat{H}_S]\rho(0)exp[is\widehat{H}_S]exp[i\widehat{H}_o(\tau-s)]\\ &= exp[-i\widehat{H}_S(\tau-s)]exp[-i\widehat{H}_B(\tau-s)]\rho_B exp[-is\widehat{H}_S]\rho(0)exp[is\widehat{H}_S]exp[i\widehat{H}_B(\tau-s)]exp[i\widehat{H}_S(\tau-s)]\\ &= exp[-i\widehat{H}_B(\tau-s)]\rho_B exp[i\widehat{H}_B(\tau-s)]exp[-i\widehat{H}_S(\tau-s)]exp[-is\widehat{H}_S]\rho(0)exp[is\widehat{H}_S]exp[i\widehat{H}_S(\tau-s)]\\ &= exp[-i\widehat{H}_B(\tau-s)]\rho_B exp[i\widehat{H}_B(\tau-s)]exp[-i\tau\widehat{H}_S]\rho(0)exp[i\tau\widehat{H}_S]\\ &= \rho_B(s-\tau)\rho(-\tau).\end{aligned}$$

$$(49)$$

In Eq. (49), we have used Baker-Campbell-Hausdorff formula

$$exp(\lambda\widehat{L})\psi = exp(\lambda\widehat{H})\psi exp(-\lambda\widehat{H}),\ \widehat{L}\psi = [\widehat{H},\psi]. \quad (50)$$

Equation (48) is further expanded as

$$\widehat{B} = \widehat{L}_{int} \widehat{U}_o(s,\tau) \widehat{L}_{int} \rho_B(s-\tau) \rho(-\tau)$$
$$= \widehat{L}_{int} e^{-i\widehat{L}_o(s-\tau)} [\widehat{H}_{int}, \rho_B(s-\tau)\rho(-\tau)]$$
$$= \widehat{L}_{int} e^{-i\widehat{H}_o(s-\tau)} [\widehat{H}_{int}, \rho_B(s-\tau)\rho(-\tau)] e^{i\widehat{H}_o(s-\tau)} \quad (51)$$
$$= \widehat{L}_{int} [\widehat{H}_{int}(\tau-s), \rho_B \rho(-s)]$$
$$= [\widehat{H}_{int}, [\widehat{H}_{int}(\tau-s), \rho_B \rho(-s)]].$$

Substituting Eq. (40) and take a trace over the environment, we get

$$tr_B(\widehat{B})$$
$$= \lambda^2 \sum_{i,j,k,l} tr_B(b_i^j b_k^l(\tau-s)\rho_B) \{ S_i^j S_k^l(\tau-s)\rho(-s) - S_k^l(\tau-s)\rho(-s) S_i^j \}$$
$$+ \lambda^2 \sum_{i,j,k,l} tr_B(b_k^l(\tau-s) b_i^j \rho_B) \{ \rho(-s) S_k^l(\tau-s) S_i^j - S_i^j \rho(-s) S_k^l(\tau-s) \}$$
$$= \lambda^2 \sum_{i,j,k,l} tr_B(b_i^j b_k^l(\tau-s)\rho_B) [S_i^j, S_k^l(\tau-s)\rho(-s)]$$
$$+ \lambda^2 \sum_{i,j,k,l} tr_B(b_k^l(\tau-s) b_i^j \rho_B) [\rho(-s) S_k^l(\tau-s), S_i^j]$$
$$= \sum_{i,j} \{ [S_i^j, S_k^l(\tau-s)\rho(-s)](\Gamma(\tau-s) - i\Delta(\tau-s)) + [\rho(-s) S_k^l(\tau-s), S_i^j](\Gamma(\tau-s) + i\Delta(\tau-s)) \}.$$
$$(52)$$

Then, we obtain

$$\widehat{V}^{(2)}(t)\rho(0)$$
$$= \widehat{U}_S(t)\rho(0) - \widehat{U}_S(t) \int_0^t ds \int_0^s d\tau \widehat{U}_S^{-1}(s) \sum_{i,j} [S_i^j, S_i^j(\tau-s)\rho(-s)](\Gamma(\tau-s) - i\Delta(\tau-s))$$
$$- \widehat{U}_S(t) \int_0^t ds \int_0^s d\tau \widehat{U}_S^{-1}(s) \sum_{i,j} [\rho(-s) S_i^j(\tau-s), S_i^j](\Gamma(\tau-s) + i\Delta(\tau-s)).$$
$$(53)$$

If we evaluate $\widehat{V}^{(2)}(t)$ in the multiplet basis, we obtain
$$\widehat{V}^{(2)}{}_{ab|cd} = (e_{ab}, \widehat{V}^{(2)}(t) e_{cd}). \quad (54)$$

Let's define operator $\widehat{\Xi}_1, \widehat{\Xi}_2$ as follows:
$$\widehat{\Xi}_1 = \widehat{U}_S^{-1}(s) \sum_{i,j} [S_i^j, S_i^j(\tau-s)\rho(-s)] = \sum_{i,j} [S_i^j(s), S_i^j(\tau)\rho(0)],$$
$$\widehat{\Xi}_2 = \widehat{U}_S^{-1}(s) \sum_{i,j} [\rho(-s) S_i^j(\tau-s), S_i^j] = \sum_{i,j} [\rho(0) S_i^j(\tau), S_i^j(s)].$$

We first calculate $(e_{ab}, \widehat{\Xi}_1)$ which becomes

$$(e_{ab}, \widehat{\Xi}_1)$$
$$= \left( e_{ab}, \sum_{i,j} \sum_{c,d} [S_i^j(s), S_i^j(\tau) e_{cd}] \rho_{cd}(0) \right) \quad (55)$$
$$= \sum_{cd} \left\{ \delta_{ab} \sum_{a'} M_{aa'a'c} e^{i\tau(E_{a'} - E_c) + is(E_a - E_{a'})} - M_{acdb} e^{i\tau(E_a - E_c) + is(E_d - E_b)} \right\}.$$

Likewise, $(e_{ab}, \widehat{\Xi}_2)$ becomes

$$\left(e_{ab}, \widehat{\Xi}_2\right)$$
$$= \left(e_{ab}, \sum_{i,j}\sum_{c,d}\left[e_{cd}S_i^j(\tau), S_i^j(s)\right]\rho_{cd}(0)\right) \qquad (55)$$
$$= \sum_{cd}\left\{\delta_{ac}\sum_{a'} M_{da'a'b}e^{i\tau(E_d-E_{a'})+is(E_{a'}-E_b)} - M_{acdb}e^{i\tau(Ed-Eb)+is(E_a-E_c)}\right\}.$$

where

$$M_{abcd} = \sum_{i,j}<a|S_i^j|b><c|S_i^j|d>$$
$$= \frac{1}{4}\sum_{i=1,2}\left\{<a|X_i|b><c|X_i|d> + <a|Y_i|b><c|Y_i|d> + <a|Z_i|b><c|Z_i|d>\right\}. \qquad (56)$$

We define $K_{ab|cd}(s,\tau)$ as

$$K_{ab|cd}(s,\tau)$$
$$= \left\{\delta_{ab}\sum_{a'} M_{aa'a'c}e^{i\tau(E_{a'}-E_c)+is(E_a-E_{a'})} - M_{acdb}e^{i\tau(E_a-E_c)+is(E_d-E_b)}\right\}(\Gamma(\tau-s) - i\Lambda(\tau-s))$$
$$+ \left\{\delta_{ac}\sum_{a'} M_{da'a'b}e^{i\tau(E_d-E_{a'})+is(E_{a'}-E_b)} - M_{acdb}e^{i\tau(Ed-Eb)+is(E_a-E_c)}\right\}(\Gamma(\tau-s) + i\Lambda(\tau-s)).$$
$$(57)$$

By integrating Eq. (57) over $ds, d\tau$, we obtain

$$\int_0^t ds \int_0^s d\tau K_{ab|cd}(s,\tau)$$
$$= \delta_{ab}\sum_{a'} M_{aa'a'c}k_{a'a'|ca}(t) - M_{acdb}k_{ab|cd}(t) + \delta_{ac}\sum_{a'} M_{da'a'b}k^*_{a'a'|db}(t) M_{acdb}k^*_{ba|dc}(t), \qquad (58)$$
$$k_{ab|cd}(t) = \int_0^t ds\, e^{is(\omega_{ac}-\omega_{bd})}\int_0^s d\tau\, e^{-i\tau\varpi_{ac}}(\Gamma(\tau) + i\Delta(\tau)),$$
$$\omega_{ab} = E_a - E_b.$$

Equation (42) becomes

$$\Gamma(t) + i\Delta(t) = \frac{\lambda^2}{\pi}\int_0^\infty J(\omega)\left\{\exp(-i\omega t) + \coth\left(\frac{\omega}{2k_BT}\right)\cos\omega t\right\}$$
$$= \frac{2\lambda^2\eta k_B T}{\pi}\frac{\sin(\omega_c t)}{t} - i\frac{\lambda^2\eta\omega_c}{\pi}\left[\frac{\sin(\omega_c t)}{\omega_c t^2} - \frac{\cos(\omega_c t)}{t}\right] \qquad (59)$$
$$= \frac{\pi\Gamma_o}{2\tau_s}\frac{\sin(\omega_c t)}{t} - i\frac{\Delta_o}{\tau_s}\left[\frac{\sin(\omega_c t)}{\omega_c t^2} - \frac{\cos(\omega_c t)}{t}\right].$$

In order to evaluate the integrals in (58) and (59), we need to calculate the integral

$$I_2(s) = \int_0^s d\tau \frac{\sin(\omega'_c \tau)}{\tau}, \text{ which is given by}$$

$$I_2(s) = \frac{\pi \omega'_c}{2} - \sin(\omega'_c s)\int_0^\infty dx \frac{xe^{-xs}}{x^2+\omega'^2_c} - \omega'_c \cos(\omega'_c s)\int_0^\infty dx \frac{e^{-xs}}{x^2+\omega'^2_c}$$

$$= \frac{\pi \omega'_c}{2} + Si(\omega'_c s), \tag{60}$$

$$\omega'_c = \omega_c \tau_s.$$

In Eq. (60), we have used the following relations [6]:

$$\int_0^\infty dx \frac{xe^{-x\mu}}{x^2+\beta^2} = -Ci(\beta\mu)\cos(\beta\mu) - Si(\beta\mu)\sin(\beta\mu),$$

$$\int_0^\infty dx \frac{e^{-x\mu}}{x^2+\beta^2} = \frac{1}{\beta}\left[Ci(\beta\mu)\sin(\beta\mu) - Si(\beta\mu)\cos(\beta\mu)\right]. \tag{61}$$

We now define $I_1(t) = \int_0^t ds\, I_2(s)$, which is given by

$$I_1(t) = \frac{\pi}{2}\omega'_c t + \int_0^t Si(\omega'_c s)\, ds. \tag{62}$$

From Eqs. (58)-(62), we obtain $k_{ab|cd}(t) = \int_0^t ds \int_0^s d\tau K_{ab|cd}(s,\tau)$ given by

$$\begin{aligned}k_{ab|cd}(t) &= \int_0^t ds \int_0^s d\tau K_{ab|cd}(s,\tau) \\ &= \frac{2}{\pi}\Gamma_o I_1\left(\frac{t}{\tau_s}\right) + i\Delta_o\left\{\frac{t}{\tau_s} - \frac{1}{\omega'_c}I_2\left(\frac{t}{\tau_s}\right)\right\} \\ &= \frac{2}{\pi}\Gamma_o\left\{\frac{\pi}{2}\omega_c t + \int_0^{\frac{t}{\tau_s}} Si(\omega_c \tau_s t)\, dt\right\} \\ &\quad + i\Delta_o\left\{\frac{t}{\tau_s} - \frac{1}{\omega_c \tau_s}\left[\frac{\pi}{2}\omega_c \tau_s + Si(\omega_c t)\right]\right\} = k(t). \end{aligned} \tag{63}$$

This establishes Eqs. (45) to Eq. (47).

### III. Two-qubit gate operations with non-Markovian noise sources
### SWAP operation

We first consider the SWAP gate operation for various input states.

We now consider the non-Markovian error associated with SWAP gate operation. For SWAP gate operation, the multiplet basis is given by

$$|1>^m = |00>,$$
$$|2>^m = \frac{1}{\sqrt{2}}(|01> + |10>),$$
$$|3>^m = |11>, \tag{64}$$
$$|4>^m = \frac{1}{\sqrt{2}}(|01> - |10>)$$

In NISQ machines, the input and out states are represented by the computational basis:
$$e^c_{\alpha\beta} = |\alpha><\beta|, \quad \alpha,\beta = 1,2,3,4, \quad \{|00>,|01>,|10>,|11>\}. \tag{65}$$

Then the reduce-density-operator in the computational basis is given by
$$\rho^c_{\alpha\beta}(t) = \sum_{a',b'} C_{\alpha\beta|a'b'} \rho^m_{a'b'}(t),$$
$$C_{\alpha\beta|ab} = tr\left(e^{c\dagger}_{\alpha\beta}, e^m_{ab}\right). \tag{66}$$

Case 1: When that the initial state is given by $\rho_{cd}(0) = |00><00| = \rho^{SWAP}_{11}(0)$, we obtain
$$\rho^{SWAP}_{11}(t) = V_{11|11}(t)\rho^{SWAP}_{11}(0)$$
$$= \left\{1 - 2\left[\sum_{a'} M_{1a'a'1} - M_{1111}\right] Re\, k(t)\right\}\rho^{SWAP}_{11}(0)$$
$$= (1 - 2\, Re\, k(t))\rho^{SWAP}_{11}(0),$$
$$\rho^{SWAP}_{22}(t) = V_{22|11}(t)\rho^{SWAP}_{11}(0)$$
$$= 2M_{2112}\, Re\, k(t)\rho^{SWAP}_{11}(0)$$
$$= Re\, k(t)\rho^{SWAP}_{11}(0),$$
$$\rho^{SWAP}_{33}(t) = V_{33|11}(t)\rho^{SWAP}_{11}(0) \tag{67}$$
$$= 2M_{3113}\, Re\, k(t)\rho^{SWAP}_{11}(0)$$
$$= 0,$$
$$\rho^{SWAP}_{44}(t) = V_{44|11}(t)\rho^{SWAP}_{11}(0)$$
$$= 2M_{4114}\, Re\, k(t)\rho^{SWAP}_{11}(0)$$
$$= Re\, k(t)\rho^{SWAP}_{11}(0).$$

Using Eqs. (64)-(66) and (67), we obtain the computational basis representation of the reduced-density-operator as

$$\rho^c_{11}(t) = (1 - 2\,Re\,k(t))\rho^{SWAP}_{11}(0),$$

$$\rho^c_{22}(t) = Re\,k(t)\rho^{SWAP}_{11}(0),$$

$$\rho^c_{33}(t) = Re\,k(t)\rho^{SWAP}_{11}(0) \tag{68}$$

$$\rho^c_{44}(t) = 0.$$

Case 2: When that the initial state is given by

$$\rho_{cd}(0) = \frac{1}{2}(|01\rangle + |10\rangle)(\langle 01| + \langle 10|) = \rho^{SWAP}_{22}(0),\ \text{we obtain}$$

$$\begin{aligned}
\rho^{SWAP}_{11}(t) &= V_{11|22}(t)\rho^{SWAP}_{22}(0) \\
&= 2M_{1221}\,Re\,k(t)\rho^{SWAP}_{22}(0) \\
&= Re\,k(t)\rho^{SWAP}_{22}(0),
\end{aligned}$$

$$\begin{aligned}
\rho^{CNOT}_{22}(t) &= V_{22|22}(t)\rho^{SWAP}_{22}(0) \\
&= \left\{1 - 2\left[\sum_{a'}M_{2a'a'2} - M_{2222}\right]Re\,k(t)\right\}\rho^{SWAP}_{22}(0) \\
&= (1 - 3\,Re\,k(t))\rho^{SWAP}_{22}(0),
\end{aligned}$$

$$\begin{aligned}
\rho^{SWAP}_{33}(t) &= V_{33|22}(t)\rho^{SWAP}_{22}(0) \\
&= 2M_{3223}\,Re\,k(t)\rho^{SWAP}_{22}(0) \\
&= Re\,k(t)\rho^{SWAP}_{22}(0),
\end{aligned} \tag{69}$$

$$\begin{aligned}
\rho^{SWAP}_{44}(t) &= V_{44|22}(t)\rho^{SWAP}_{22}(0) \\
&= 2M_{4224}\,Re\,k(t)\rho^{SWAP}_{22}(0) \\
&= Re\,k(t)\rho^{SWAP}_{22}(0).
\end{aligned}$$

Using Eqs. (64)-(67) and (69), we obtain the computational basis representation of the reduced-density-operator as

$$\rho^c_{11}(t) = Re\,k(t)\rho^{SWAP}_{22}(0),$$

$$\rho^c_{22}(t) = \frac{1}{2}(1 - 2\,Re\,k(t))\rho^{SWAP}_{22}(0),$$

$$\rho^c_{33}(t) = \frac{1}{2}(1 - 2\,Re\,k(t))\rho^{SWAP}_{22}(0) \tag{70}$$

$$\rho^c_{44}(t) = Re\,k(t)\rho^{SWAP}_{22}(0).$$

Case 3: When that the initial state is given by $\rho_{cd}(0) = |11\rangle\langle 11| = \rho^{SWAP}_{33}(0)$, we obtain

$$\rho_{11}^{SWAP}(t) = V_{11|33}(t)\rho_{33}^{SWAP}(0)$$
$$= 2M_{1331} Re\, k(t)\rho_{33}^{SWAP}(0)$$
$$= 0,$$
$$\rho_{22}^{SWAP}(t) = V_{22|33}(t)\rho_{33}^{SWAP}(0)$$
$$= 2M_{2332} Re\, k(t)\rho_{33}^{SWAP}(0)$$
$$= Re\, k(t)\rho_{33}^{SWAP}(0),$$
$$\rho_{33}^{SWAP}(t) = V_{33|33}(t)\rho_{33}^{SWAP}(0) \qquad (71)$$
$$= \left\{1 - 2\left[\sum_{a'} M_{3a'a'3} - M_{3333}\right] Re\, k(t)\right\}\rho_{33}^{SWAP}(0)$$
$$= (1 - 2Re\, k(t))\rho_{33}^{SWAP}(0),$$
$$\rho_{44}^{SWAP}(t) = V_{44|33}(t)\rho_{33}^{SWAP}(0)$$
$$= 2M_{4334} Re\, k(t)\rho_{33}^{SWAP}(0)$$
$$= Re\, k(t)\rho_{33}^{SWAP}(0).$$

Using Eqs. (64)-(67) and (71), we obtain the computational basis representation of the reduced-density-operator as

$$\rho_{11}^c(t) = 0,$$
$$\rho_{22}^c(t) = Re\, k(t)\rho_{33}^{SWAP}(0),$$
$$\rho_{33}^c(t) = Re\, k(t)\rho_{33}^{SWAP}(0), \qquad (72)$$
$$\rho_{44}^c(t) = (1 - Re\, k(t))\rho_{33}^{SWAP}(0).$$

Case 4: When that the initial state is given by

$$\rho_{cd}(0) = \frac{1}{2}(|01\rangle - |10\rangle)(\langle 01| - \langle 10|) = \rho_{44}^{SWAP}(0),$$ we obtain

$$\rho_{11}^{SWAP}(t) = V_{11|44}(t)\rho_{44}^{SWAP}(0)$$
$$= 2M_{1441} Re\, k(t) \rho_{44}^{SWAP}(0)$$
$$= Re\, k(t) \rho_{44}^{SWAP}(0),$$

$$\rho_{22}^{SWAP}(t) = V_{22|44}(t)\rho_{44}^{SWAP}(0)$$
$$= 2M_{2442} Re\, k(t) \rho_{44}^{SWAP}(0)$$
$$= Re\, k(t) \rho_{44}^{SWAP}(0),$$

$$\rho_{33}^{SWAP}(t) = V_{33|44}(t)\rho_{44}^{SWAP}(0) \quad (73)$$
$$= 2M_{3443} Re\, k(t) \rho_{44}^{SWAP}(0)$$
$$= Re\, k(t) \rho_{44}^{SWAP}(0),$$

$$\rho_{44}^{SWAP}(t) = V_{44|44}(t)\rho_{44}^{SWAP}(0)$$
$$= \left\{ 1 - 2\left[\sum_{a'} M_{4a'a'4} - M_{4444}\right] Re\, k(t) \right\} \rho_{44}^{SWAP}(0)$$
$$= (1 - 3Re\, k(t)) \rho_{44}^{SWAP}(0).$$

Using Eqs. (66)-(67) and (73), we obtain the computational basis representation of the reduced-density-operator as

$$\rho_{11}^c(t) = Re\, k(t) \rho_{44}^{SWAP}(0),$$
$$\rho_{22}^c(t) = \frac{1}{2}(1 - 2Re\, k(t)) \rho_{44}^{SWAP}(0),$$
$$\rho_{33}^c(t) = \frac{1}{2}(1 - 2Re\, k(t)) \rho_{44}^{SWAP}(0), \quad (74)$$
$$\rho_{44}^c(t) = Re\, k(t) \rho_{44}^{SWAP}(0).$$

**Identity operation**

We now consider the non-Markovian error associated with Identity gate operation. For Identity gate operation, the multiplet basis is the same as the computational basis given by

$$|1\rangle = |00\rangle,$$
$$|2\rangle = |01\rangle, \quad (75)$$
$$|3\rangle = |10\rangle,$$
$$|4\rangle = |11\rangle.$$

Case 1: When that the initial state is given by $\rho_{cd}(0) = |00\rangle\langle 00| = \rho_{11}^{Id}(0)$, we obtain

$$\begin{aligned}
\rho_{11}^{Id}(t) &= V_{11|11}(t)\rho_{11}^{Id}(0) \\
&= \left\{1 - 2\left[\sum_{a'} M_{1a'a'1} - M_{1111}\right] \operatorname{Re} k(t)\right\} \rho_{11}^{Id}(0) \\
&= (1 - 2\operatorname{Re} k(t))\rho_{11}^{Id}(0), \\
\rho_{22}^{Id}(t) &= V_{22|11}(t)\rho_{11}^{Id}(0) \\
&= 2M_{2112}\operatorname{Re} k(t)\rho_{11}^{Id}(0) \\
&= \operatorname{Re} k(t)\rho_{11}^{Id}(0), \\
\rho_{33}^{Id}(t) &= V_{33|11}(t)\rho_{11}^{Id}(0) \\
&= 2M_{3113}\operatorname{Re} k(t)\rho_{11}^{Id}(0) \\
&= \operatorname{Re} k(t)\rho_{11}^{Id}(0), \\
\rho_{44}^{SWAP}(t) &= V_{44|11}(t)\rho_{11}^{Id}(0) \\
&= 2M_{4114}\operatorname{Re} k(t)\rho_{11}^{Id}(0) \\
&= 0.
\end{aligned} \qquad (76)$$

Case 2: When that the initial state is given by $\rho_{cd}(0) = |01\rangle\langle01| = \rho_{22}^{Id}(0)$, we obtain

$$\begin{aligned}
\rho_{11}^{Id}(t) &= V_{11|22}(t)\rho_{22}^{Id}(0) \\
&= 2M_{1221}\operatorname{Re} k(t)\rho_{22}^{Id}(0) \\
&= \operatorname{Re} k(t)\rho_{22}^{Id}(0), \\
\rho_{22}^{CNOT}(t) &= V_{22|22}(t)\rho_{22}^{Id}(0) \\
&= \left\{1 - 2\left[\sum_{a'} M_{2a'a'2} - M_{2222}\right] \operatorname{Re} k(t)\right\} \rho_{22}^{Id}(0) \\
&= (1 - 2\operatorname{Re} k(t))\rho_{22}^{Id}(0), \\
\rho_{33}^{Id}(t) &= V_{33|22}(t)\rho_{22}^{Id}(0) \\
&= 2M_{3223}\operatorname{Re} k(t)\rho_{22}^{Id}(0) \\
&= 0, \\
\rho_{44}^{id}(t) &= V_{44|22}(t)\rho_{22}^{Id}(0) \\
&= 2M_{4224}\operatorname{Re} k(t)\rho_{22}^{Id}(0) \\
&= \operatorname{Re} k(t)\rho_{22}^{Id}(0).
\end{aligned} \qquad (77)$$

Case 3: When that the initial state is given by $\rho_{cd}(0) = |10\rangle\langle10| = \rho_{33}^{Id}(0)$, we obtain

$$\rho_{11}^{Id}(t) = V_{11|33}(t)\rho_{33}^{Id}(0)$$
$$= 2M_{1331} Re\, k(t)\rho_{33}^{Id}(0)$$
$$= Re\, k(t)\rho_{33}^{Id}(0),$$
$$\rho_{22}^{Id}(t) = V_{22|33}(t)\rho_{33}^{Id}(0)$$
$$= 2M_{2332} Re\, k(t)\rho_{33}^{Id}(0)$$
$$= 0,$$
$$\rho_{33}^{Id}(t) = V_{33|33}(t)\rho_{33}^{Id}(0) \quad (78)$$
$$= \left\{1 - 2\left[\sum_{a'} M_{3a'a'3} - M_{3333}\right]Re\, k(t)\right\}\rho_{33}^{Id}(0)$$
$$= (1 - 2Re\, k(t))\rho_{33}^{Id}(0),$$
$$\rho_{44}^{Id}(t) = V_{44|33}(t)\rho_{33}^{Id}(0)$$
$$= 2M_{4334} Re\, k(t)\rho_{33}^{Id}(0)$$
$$= Re\, k(t)\rho_{33}^{Id}(0).$$

Case 4: When that the initial state is given by $\rho_{cd}(0) = |11\rangle\langle 11| = \rho_{44}^{Id}(0)$, we obtain

$$\rho_{11}^{Id}(t) = V_{11|44}(t)\rho_{44}^{Id}(0)$$
$$= 2M_{1441} Re\, k(t)\rho_{44}^{Id}(0)$$
$$= 0,$$
$$\rho_{22}^{Id}(t) = V_{22|44}(t)\rho_{44}^{Id}(0)$$
$$= 2M_{2442} Re\, k(t)\rho_{44}^{Id}(0)$$
$$= Re\, k(t)\rho_{44}^{Id}(0),$$
$$\rho_{33}^{Id}(t) = V_{33|44}(t)\rho_{44}^{Id}(0) \quad (79)$$
$$= 2M_{3443} Re\, k(t)\rho_{44}^{Id}(0)$$
$$= Re\, k(t)\rho_{44}^{Id}(0),$$

$$\rho_{44}^{Id}(t) = V_{44|44}(t)\rho_{44}^{Id}(0)$$
$$= \left\{1 - 2\left[\sum_{a'} M_{4a'a'4} - M_{4444}\right]Re\, k(t)\right\}\rho_{44}^{Id}(0)$$
$$= (1 - 2Re\, k(t))\rho_{44}^{Id}(0).$$

We have tried both ibm_guadalupe through IBM Quantum and IonQ through Amazon Braket to compare the theory with the experiment for the non-Markovian errors associated with SWAP and Identity operations.